\documentclass[aps,prf,superscriptaddress]{revtex4-2}

\usepackage{graphicx} 
\usepackage[dvipsnames]{xcolor}
\usepackage{amsmath} 
\usepackage[normalem]{ulem}

\title{HIT_enstrophy_pr}

\begin{document}

\title{\textbf{Two-point enstrophy dynamics in homogeneous isotropic turbulence} 
}%

\author{Gabriele Boga}
 \email{Contact author: gabriele.boga@unimore.it}
 \altaffiliation[]{DIEF, University of Modena and Reggio Emilia, 41125 Modena, Italy.}
\affiliation{%
  DIEF, University of Modena and Reggio Emilia, 41125 Modena, Italy
}%

\author{Carlos B. da Silva}
\affiliation{
LAETA, IDMEC, Instituto Superior Técnico,\\
 Universidade de Lisboa, Av. Rovisco Pais, 1049-001 Lisbon, Portugal
}%
\author{Sergio Chibbaro}
\affiliation{%
 Université Paris-Saclay, CNRS,  UMR 9015, LISN, F-91405, Orsay Cedex, France
}%

\author{Andrea Cimarelli}%
\affiliation{%
  DIEF, University of Modena and Reggio Emilia, 41125 Modena, Italy
}%

\date{\today}

\begin{abstract}
In the present work we investigate the multiscale dynamics of enstrophy in homogeneous isotropic turbulence by exploiting the two-point formalism provided by the Kármán–Howarth–Monin–Hill approach extended to the enstrophy governing equation. The study is conducted on direct numerical simulations with a Taylor-based Reynolds number in the range of $140 \lesssim Re_{\lambda} \lesssim 400$. The two-point enstrophy budget at scales $r > 10 \eta$ appears to be entirely determined by production via vortex stretching, which balances enstrophy destruction, and to be dominated by the diffusive transport at smaller scales, thus preventing the emergence of a range dominated by the inertial transport of enstrophy. Consequently, the second-order structure function of vorticity only shows a viscous scaling $\sim r^2$ at very small scales, followed by a constant behavior at larger scales. This plateau, together with the statistical isotropy, implies that the structure function is equally distributed among its longitudinal and transverse contributions at such scales. The decomposition in longitudinal and transverse contributions also highlights a dual nature of the inertial enstrophy flux. In particular, enstrophy appears to be transferred across scales through a non-trivial combination of direct and reverse interscale transfer. It is shown that the dual nature of this transfer is strictly related to the vortex stretching mechanism, which, in addition to producing enstrophy through vorticity amplification, also transfers longitudinal vorticity towards larger scales (by stretching the vortical elements) and transverse vorticity towards smaller scales (by contracting these vortical elements in the radial direction). The sum of these two contributions results in an overall transfer of enstrophy from large towards small scales. We propose the use of the pressure transport term as a proxy to obtain some information on the dynamics of relevant events of inertial energy and enstrophy transport. The new findings highlight the relevance of inertial compression events in longitudinal energy transport. At the same time, a good correlation between transverse energy transport events and the radial contraction of vortical elements due to vortex stretching mechanisms is also found.
\end{abstract}


\maketitle

\section{Introduction}
The cornerstone of turbulent dynamics is the turbulent cascade phenomenon, a nonequilibrium statistical process through which conserved quantities are transferred across a range of scales~\cite{kolmogorov41a,kolmogorov41b,kolmogorov41c,frisch,dubrulle}. This original process is key in many applications~\cite{alexakis}.
In three-dimensional turbulence, the dominant mechanism is the forward cascade of kinetic energy from large to small scales. Although the statistical signatures of this process were established in Kolmogorov’s seminal works~\cite{kolmogorov41a,kolmogorov41b,kolmogorov41c}, notably through the skewness of velocity increments, the detailed dynamical mechanisms that sustain the energy cascade mechanism remain comparatively poorly understood.

Since the early argument of Taylor~\cite{taylor38}, and following the implications of Kelvin’s theorem, vortex stretching has long been regarded as the fundamental engine driving the forward energy cascade~\cite{onsager1949statistical,Tennekes}. Scaling arguments within the Reynolds-Avergaed Navier-Stokes (RANS) framework, together with physically reasonable assumptions, support the view that, in turbulent flows, the three-dimensional Navier–Stokes equations generate a net production of enstrophy~\cite{Tennekes}. Moreover, on average, strain amplification may be expressed in terms of vortex stretching thanks to the Betchov relation~\cite{Betchov1956}.
More recent analyses have revisited Taylor’s argument using modern mathematical tools, confirming its validity in at least a statistical sense~\cite{constantin2008stochastic,eyink}.
Thus, even though direct numerical simulations and scale-by-scale analysis have shown that the Taylor picture is a bit too simplistic~\cite{Tsinober_Kit_Dracos_1992,tsinober2009informal,CarboneBragg2020JFM,Johnson2021JFM}, the understanding of the turbulent cascade remains linked to the vorticity and enstrophy dynamics~\cite{kerr1985,buaria2020vortex}, and vortex stretching remains a crucial mechanism for the energy cascade.

Despite this classical picture, a detailed multiscale understanding of the vorticity dynamics underlying the forward energy cascade remains incomplete. Prior work has predominantly focused on the energy cascade, whereas the scale-by-scale structure of vorticity and enstrophy transfer has been much less explored. In particular, coarse-grained analyses predict that vortex stretching should dominate the enstrophy budget throughout the inertial range~\cite{Tennekes}. Relatively recent developments
are presented in the pioneering work~\cite{Davidson2008}, where the authors investigate the scale-by-scale enstrophy dynamics in homogeneous isotropic turbulence using the filtering formalism. The results highlight the dual nature of the vortex stretching phenomenon, which is simultaneously responsible for a interscale enstrophy flux towards smaller scales and a generation of enstrophy through vorticity amplification. The same formalism is adopted in \cite{Doan2018}, where the authors provides interesting insights on the interaction between different scales in the vortex stretching phenomenon. More recently, the two-point formalism has been used to investigate the multiscale enstrophy dynamics both in classical homogeneous isotropic turbulence \cite{Simultaneous_casc_2022} and in polymeric turbulence \cite{Chiarini_2025}. However, a full understanding of the multiscale enstrophy dynamics and of its connection with the energy cascade is still lacking.
The present work complements the previous works filling this gap through a systematic investigation of the scale-by-scale enstrophy budget using high-resolution direct numerical simulations of isotropic turbulence. 

Analyses of scale-by-scale budgets have already proven to provide many insights on different aspects of turbulent flows~\cite{Johnson_annurev,yao2024comparing}. Specifically, the present work uses the Kármán–Howarth–Monin–Hill approach~\cite{Hill2002exact,marati2004energy,danaila2012yaglom,Cimarelli_Boga_Pavan_Costa_Stalio_2024,Gatti_Chiarini_Cimarelli_Quadrio_2020,vassilicos91}, also known as the Generalized Kolmogorov Equation (GKE) framework. By evaluating the contributions of vortex stretching, nonlinear transport, and viscous effects in the two-point enstrophy equation, we quantify their respective roles across scales and analyze their connection to the forward energy cascade. In particular, the results of the present work are aimed at clarifying the multiscale structure of vorticity dynamics and the dynamical foundation of the turbulent cascade mechanism. A further insight in the multiscale enstrophy dynamics is provided by decomposing the interscale enstrophy flux in its longitudinal and transverse contributions. This decomposition is also used to propose a mechanistic interpretation of the enstrophy flux and to attempt to establish a link with the energy cascade process.

The paper is organized as follows. Section \ref{th_bckgrnd} provides the two-point formalism and the relative enstrophy budget equations used in the present work. The results are presented in section \ref{results}. The investigation begins with the analysis of the two-point enstrophy budget in subsection \ref{twopoint_enstr} and proceeds with the study of the second-order vorticity structure function and its longitudinal and transverse contributions in subsection \ref{struct_func}. Subsection \ref{enstr_inert_flux} analyzes the inertial enstrophy flux and provides an interpretation of its underlying phenomenology. Finally, in subsection \ref{pressure}, conditional correlations between different transport terms are used both to provide support to the scenario depicted in the previous subsection (\ref{enstr_inert_flux}) and to link it to the energy cascade phenomenology by using the pressure transport term as a proxy. The paper is closed with concluding remarks in section \ref{concl}. Appendix \ref{dataset} details the numerical simulations, while in Appendix \ref{energy} we report the results on the two-point energy budget as a reference for the less-known enstrophy budget analyzed.

\section{Theoretical background} \label{th_bckgrnd}
Enstrophy, defined as $\zeta = \omega_i \omega_i$ (where $\omega_i = \epsilon_{ijk} \partial u_k / \partial x_j$ is the vorticity vector and $\epsilon_{ijk}$ is the permutation tensor), is a quantity of interest in turbulent flows as it may give an alternative insight into what can be observed through energy-related quantities. In fact, enstrophy is related to both the energy cascade process through the vortex stretching mechanism, together with the strain self-amplification~\citep{Johnson2020PRL,Johnson2021JFM}, and with the viscous dissipation rate of kinetic energy.

The latter is defined as $\varepsilon = 2 \nu s_{ij} s_{ij}$, where $s_{ij} = (\partial u_i / \partial x_j + \partial u_j / \partial x_i)/2$ and $\nu$ is the kinematic viscosity. In the following, we refer to $\varepsilon$ simply as the dissipation. It is also useful to introduce the pseudo-dissipation, defined as $\varepsilon^* = \nu (\partial u_i / \partial x_j) (\partial u_i / \partial x_j)$. Both quantities can be related to the enstrophy $\zeta$ through the relation:
\begin{equation}
\begin{gathered}
\varepsilon = \nu \zeta + 2 \nu \frac{\partial u_i}{\partial x_j} \frac{\partial u_j}{\partial x_i}
\qquad \text{and} \qquad \varepsilon^* = \nu \zeta + \nu \frac{\partial u_i}{\partial x_j} \frac{\partial u_j}{\partial x_i}
\end{gathered}
\end{equation}

Under statistically homogeneous and isotropic conditions,
\begin{equation}
\langle \frac{\partial u_i}{\partial x_j} \frac{\partial u_j}{\partial x_i} \rangle = \langle \frac{\partial^2 u_i u_j}{\partial x_i \partial x_j} \rangle = \frac{\partial^2 \langle u_i u_j \rangle}{\partial x_i \partial x_j} = 0 \\
\end{equation}
where the angular brackets $\langle \cdot \rangle$ represent spatial average operator taken over the whole physical domain. It follows that, on average,
\begin{equation}
\langle \varepsilon \rangle = \langle \varepsilon^* \rangle = \nu \langle \zeta \rangle,
\label{eq:betchov}
\end{equation}
so that the mean dissipation and pseudo-dissipation coincide and are directly related to the mean enstrophy.

\begin{figure}[ht!]
\centering
\includegraphics[width=0.95\linewidth]{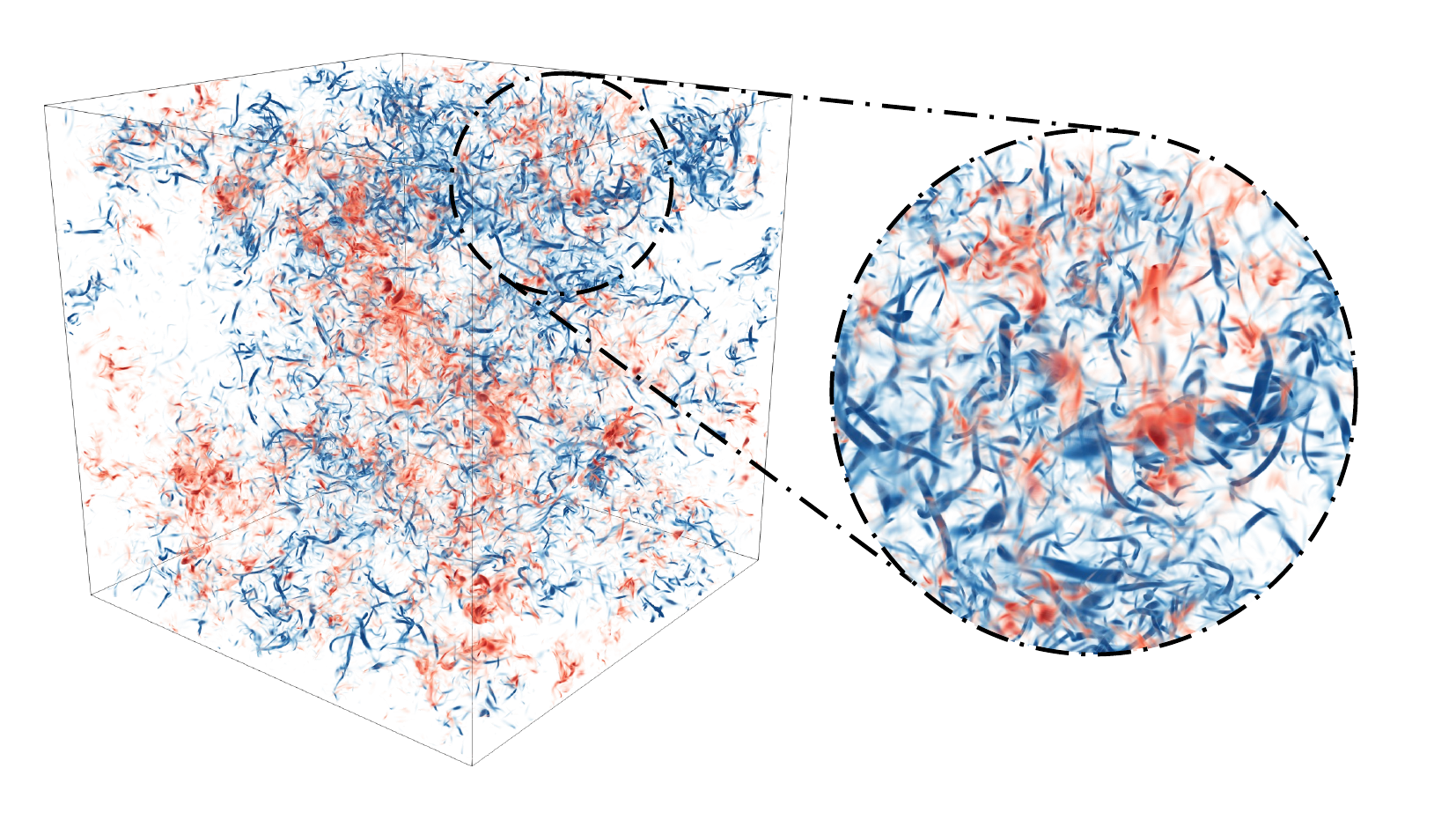}
	\caption{Instantaneous snapshot of the enstrophy $\zeta = \omega_i \omega_i$ and dissipation $\varepsilon$ fields in homogeneous isotropic turbulence at the intermediate Reynolds number simulation used in the present work ($Re_{\lambda} \approx 240$). Volume rendering of enstrophy (from transparent blue, indicating low values, to opaque blue, indicating higher values) and of dissipation (from transparent red, indicating low values, to opaque red, indicating higher values).}
\label{fig:volrender}
\end{figure}

It is important to bear in mind that, while the mean values of the rate-of-strain magnitude squared (proportional to the dissipation rate) and the enstrophy are equal, $\langle \omega_i \omega_i \rangle = 2 \langle s_{ij} s_{ij} \rangle$, their topology is different \citep{CarboneBragg2020JFM}, as shown in figure \ref{fig:volrender}.
To study the scale-dependent properties of enstrophy we introduce the second-order structure function of vorticity (i.e. two-point enstrophy), defined as $\delta \omega^2 = \delta \omega_i \delta \omega_i$, where $\delta \omega_i = \omega_i(x_j'',t) - \omega_i(x_j',t)$, and the distance between the two points defines the space of scales $\boldsymbol{r} = \boldsymbol{x'} - \boldsymbol{x''}$. Following \cite{Simultaneous_casc_2022}, the evolution equation of two-point enstrophy written in spherical coordinates for statistically steady homogeneous and isotropic conditions reads:
\begin{equation}
\begin{gathered}
\underbrace{-\frac{1}{r^2}\frac{d}{d r} \bigg ( r^2 \langle \delta \omega^2 \delta u_r \rangle \bigg )}_{\langle T_{\omega} \rangle} \underbrace{+ 2\nu \frac{1}{r^2} \frac{d}{d r} \left ( r^2 \frac{d \langle \delta \omega^2 \rangle}{d r} \right )}_{\langle D_{\omega} \rangle} + \underbrace{2 \langle \delta \omega_i \widetilde{\omega}_j \delta \left( \frac{\partial u_i}{\partial x_j} \right) \rangle + 2 \langle \delta \omega_i \delta \omega_j \widetilde{\left( \frac{\partial u_i}{\partial x_j} \right)} \rangle}_{\langle \mathcal{S} \rangle} = 4 \langle \chi \rangle,
\label{eq:iso_enstr}
\end{gathered}
\end{equation}
where the operator $\widetilde{\cdot}$ is defined as the two-point average, e.g. $\widetilde{\omega}_i = 1/2 ( \omega_i (x_j'',t) + \omega_i (x_j',t) )$. In equation (\ref{eq:iso_enstr}), the terms $T_{\omega}$ and $D_{\omega}$ represent the inertial and diffusive transports of enstrophy across the space of scales, respectively, $\mathcal{S}$ is the vortex stretching term, and $ \chi = \nu \left(\partial \omega_i / \partial x_j \right) \left(\partial \omega_i / \partial x_j \right)$ is the viscous dissipation of enstrophy associated with spatial vorticity gradients. Note that the divergence terms in the space of scales ($T_{\omega}$ and $D_{\omega}$) are written using spherical coordinates $\boldsymbol{r} = (r, \phi, \theta)$ in order to highlight the dependence of the scale-space fluxes on the sole radial direction $r$ (since isotropic conditions are verified) while the source/sink terms are written in Cartesian coordinates for the sake of simplicity. 

By integrating equation (\ref{eq:iso_enstr}) over a spherical volume in the space of scales, we obtain a formulation highlighting enstrophy fluxes:
\begin{equation}
\begin{gathered}
\langle \delta \omega^2 \delta u_r \rangle - 2\nu \frac{d \langle \delta \omega^2 \rangle}{d r}  = \frac{1}{4 \pi r^2} \int_{\mathcal{B}_r} \langle \mathcal{S} \rangle dV - \frac{4}{3} \langle \chi \rangle r,
\label{eq:iso_enstr_flux}
\end{gathered}
\end{equation}
where $\mathcal{B}_r$ is a spherical volume of radius $r$. Equations (\ref{eq:iso_enstr}) and (\ref{eq:iso_enstr_flux}) are equal to the analogous equations governing the scale-energy $\langle \delta q^2 \rangle = \langle \delta u_i \delta u_i \rangle$ described in Appendix \ref{energy}, as equations (\ref{eq:iso_gke}) and (\ref{eq:iso_flux_gke}) respectively, with the only difference consisting in the presence of the vortex stretching term, acting as a source, which does not vanish in homogeneous isotropic conditions.

The analysis is carried out on a dataset of homogeneous isotropic turbulence consisting of three simulations at different Taylor based Reynolds numbers $Re_{\lambda} \approx 140, ~240$ and $400$ (see Appendix \ref{dataset} for further details). In figure \ref{fig:volrender} we show a typical visualization of the enstrophy and dissipation, that is strain. That helps to capture at a qualitative level some features of isotropic turbulence. In particular, we find the known picture of the high-value vorticity concentrated in filaments, while the strain appears more distributed~\cite{douady1991direct,moisy2004geometry}.

\section{Results} \label{results}
\subsection{Two-point budget} \label{twopoint_enstr}
\begin{figure}[ht!]
\centering
\includegraphics[width=0.6\linewidth]{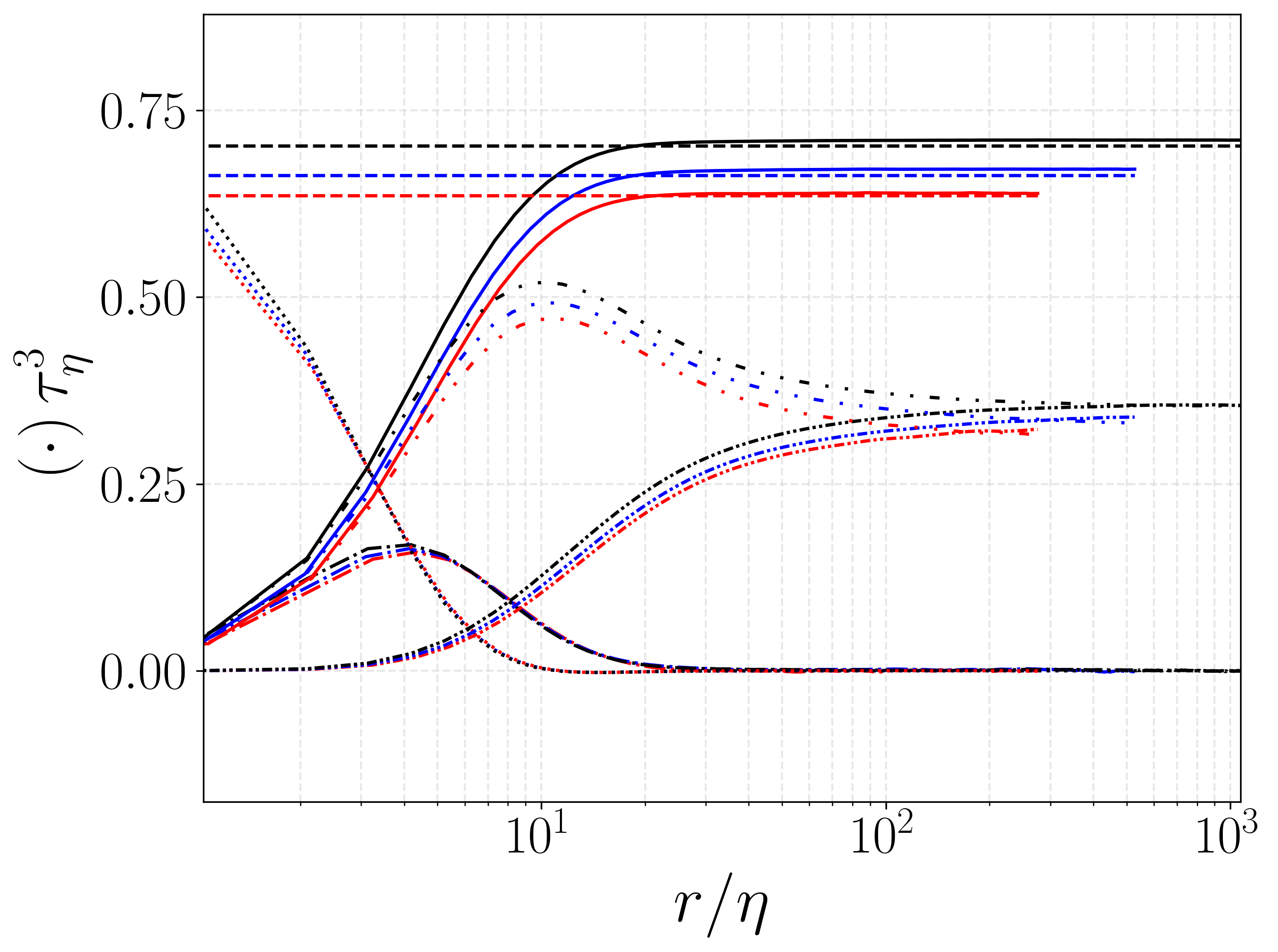}
        \caption{ Budgets of $\langle \delta \omega^2\rangle$ at Taylor based Reynolds numbers of $Re_{\lambda} \approx 140$ (red lines), $Re_{\lambda} \approx 240$ (blue lines) and $Re_{\lambda} \approx 400$ (black lines). Enstrophy destruction $4 \langle \chi \rangle$ (dashed lines), diffusive transport $\langle D_\omega \rangle$ (dotted lines), inertial transport $\langle T_\omega \rangle$ (dash-dotted lines) and vortex stretching $\langle \mathcal{S} \rangle$ (solid lines). The two vortex stretching contributions are also reported: $2 \langle \delta \omega_i \widetilde{\omega}_j \delta \left( \partial u_i / \partial x_j \right) \rangle$ (dash-dash-dotted lines) and $2 \langle \delta \omega_i \delta \omega_j \widetilde{\left( \partial u_i / \partial x_j \right) \rangle}$ (loosely dash-dash-dotted lines). All the curves are made dimensionless by using Kolmogorov units.}
\label{fig:enstr_budg}
\end{figure}

The terms of the budget equation (\ref{eq:iso_enstr}) are shown in figure \ref{fig:enstr_budg}.
At scales $r > 10 \eta$, the budget is entirely determined by production via vortex stretching $\langle S \rangle$, which balances the system’s invariant $\langle \chi \rangle$, representing enstrophy destruction through vorticity gradients:
\begin{equation}
    \langle \mathcal{S} \rangle = 4 \langle \chi \rangle,
\end{equation}
thus suggesting a local enstrophy balance in the space of scales. As in the budget of scale-energy $\langle \delta q^2 \rangle = \langle \delta u_i \delta u_i \rangle$ (figure \ref{fig:energy_budg} in Appendix \ref{energy}), the diffusive transport is dominant at small scales, although the viscous-dominated subrange is found to be narrower, with $ \langle D_\omega \rangle$ becoming negligible for $r > 10 \eta$. A major difference to the scale-energy budget lies in the inertial transport. Specifically, while in the budget of $\langle \delta q^2 \rangle$ it is possible to observe the development of a range completely dominated by inertial transport (for sufficiently high Reynolds numbers), the same is not true in the enstrophy budget. The inertial enstrophy transport always appears to be much lower than the sum of diffusive transport and vortex stretching and does not seem to increase its relevance as the Reynolds number increases. Furthermore, the peak of the inertial enstrophy transport appears to be located around $r \approx 4 \eta$, corresponding to the viscous subrange of $\langle \delta q^2 \rangle$, which may seem surprising at first. However, the inertial enstrophy transport observed at such small scales ($1 \lesssim r / \eta \lesssim 10$) should not be interpreted as a turbulent cascade process driven by inertial velocity interactions, since the velocity field can be considered approximately smooth at such scales. In contrast, enstrophy is linked to velocity derivatives and therefore involves a smaller-scale structure, becoming smooth only at even finer scales than the velocity field. Consequently, the inertial enstrophy transport at these small scales should be more appropriately interpreted as a laminar shearing phenomenon, rather than a fully turbulent transport~\cite{Batchelor_1959}. In this picture, a uniform velocity shear can produce a stirring in the non-smooth enstrophy field, bringing closer together the enstrophy isosurfaces, hence acting as an interscale enstrophy transport.

An important feature of the scale-enstrophy budget is the absence of an intermediate range of scales (with a size that increases with the Reynolds number), in between the ``enstrophy producing'' scales and the ``diffusion-dominated'' scales. Indeed, none of the terms in the budget of $\langle \delta \omega^2 \rangle$ show a scale separation increasing with the Reynolds number, in accordance with the idea that enstrophy is related to small-scale motions, which are thought to be universal. This is in strong contrast to the scale-energy budget (reported in Appendix \ref{energy}, figure \ref{fig:energy_budg}), where the ``energy injection'' scales and the ``diffusion dominated'' scales are linked through the inertial energy transport which widens as the Reynolds number increases.

The absence of a separation of scales by increasing $Re_{\lambda}$ and of a range dominated by inertial enstrophy transport, and thus of a pure enstrophy cascade, is not surprising. In fact, the vortex stretching mechanism (acting as a source in the enstrophy budget) is closely connected with the energy cascade process. Thus, it is reasonable to expect it to be active in the inertial energy subrange of scales where the inertial transport of energy is dominant, which can be confirmed by comparing figure \ref{fig:enstr_budg} and \ref{fig:energy_budg}. By definition, at scales smaller than these, diffusive transport starts to be relevant. These observations imply that, independently of the Reynolds number, there cannot be a region in which the inertial transport of enstrophy dominates with respect to both vortex stretching and diffusive transport.

Figure \ref{fig:enstr_budg} also shows the two vortex stretching contributions, namely $\langle \mathcal{S} \rangle = 2 \langle \delta \omega_i \delta \omega_j \widetilde{( \partial u_i / \partial x_j )} \rangle + 2 \langle \delta \omega_i \widetilde{\omega}_j \delta ( \partial u_i / \partial x_j ) \rangle$. The first contribution can be expressed as
\begin{equation}
2 \langle \delta \omega_i \delta \omega_j \widetilde{\left( \frac{\partial u_i}{\partial x_j} \right)} \rangle = 2 \langle \delta \omega_i \delta \omega_j \frac{\partial \delta u_i}{\partial r_j} \rangle
\end{equation}
and is the vortex stretching contribution given by the velocity increment gradients in the space of scales. The two-point increment operator $\delta \cdot$ applied to $u_j$ and $\omega_j$ has the effect of highlighting motions with a scale similar to the separation length $r$ considered \cite{Germano_twopoint}. Hence, the peak exhibited around $r \approx 10 \eta$ can be interpreted as a high vortex stretching activity around this scale, which represents a good compromise between having both a significant $\delta \omega_i$ (generally associated with small scales) and a significant $\delta u_i$ (generally associated with large scales).

On the other hand, the second contribution can be expressed as
\begin{equation}
2 \langle \delta \omega_i \widetilde{\omega}_j \delta \left( \frac{\partial u_i}{\partial x_j} \right) \rangle = 2 \langle \delta \omega_i \widetilde{\omega}_j \frac{\partial \delta u_i}{\partial x_{c_j}} \rangle
\end{equation}
and is the vortex stretching contribution given by the velocity increment gradients in the physical space (where $\boldsymbol{x_c} = (\boldsymbol{x'} + \boldsymbol{x''})/2$ is the mid-point).
The operator $\widetilde{\cdot}$ (applied to the vorticity $\omega_j$) has an effect similar to a low-pass filter, being more influenced by the scales larger than the separation length $r$ considered \cite{Germano_twopoint}. Furthermore, this term contains the gradient of the velocity increment with respect to the midpoint $\left( \partial \delta u_i / \partial x_{c_j} \right)$. Since the term $\delta u_i$ is, in a sense, more sensitive to a change in $\boldsymbol{r}$ than to a change in $\boldsymbol{x_c}$, it is not surprising that this term does not peak around a specific scale but slowly increases until reaching the single-point asymptotic value.

\subsection{Second-order structure function} \label{struct_func}

Following the observations made on the two-point enstrophy budget, it is possible to deduce the scaling laws of the structure function $\langle \delta \omega^2 \rangle$ itself. By considering very small values of $r$, within the viscous subrange ($r \approx \eta$), where equation (\ref{eq:iso_enstr_flux}) is dominated by the diffusive transport, and by integrating equation (\ref{eq:iso_enstr_flux}) in $r$, one recovers the viscous scaling law for the second-order structure function of enstrophy,
\begin{equation}
\langle \delta \omega^2 \rangle = \frac{1}{3 \nu} \langle \chi \rangle r^2,
\label{eq:viscous_enstr_scaling}
\end{equation}
which is analogous to the viscous scaling law of $\langle \delta q^2 \rangle$ reported in Appendix \ref{energy}, equation (\ref{eq:viscous_scaling}).
\begin{figure}[ht!]
\centering
\includegraphics[width=0.6\linewidth]{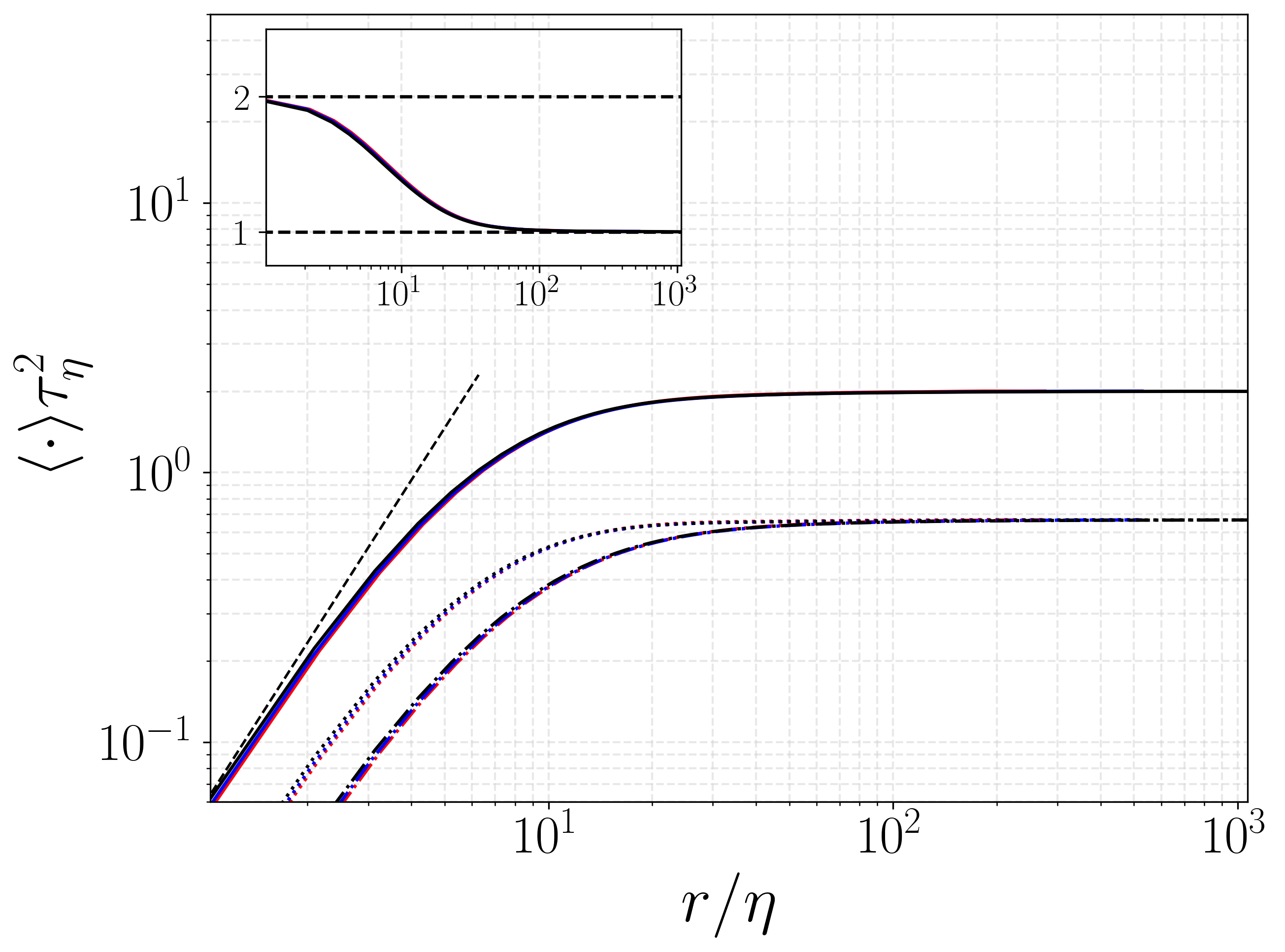}
	\caption{ Second-order structure function of enstrophy, $\langle \delta \omega^2 \rangle$ (solid lines), and its longitudinal $\langle \delta \omega_{||}^2 \rangle$ (dash-dotted lines) and transverse $\langle \delta \omega_{\perp}^2 \rangle$ (dotted lines) contributions. The viscous scaling law $\langle \delta \omega^2 \rangle = 1/(3 \nu) \langle \chi \rangle r^2$ is reported (dashed black line). The colors correspond to the three different simulations, with Taylor based Reynolds numbers of $Re_{\lambda} \approx 140$ (red lines), $Re_{\lambda} \approx 240$ (blue lines) and $Re_{\lambda} \approx 400$ (black lines). In the inset panel, the ratio $\langle \delta \omega_{\perp}^2 \rangle / \langle \delta \omega_{||}^2 \rangle$ is reported together with its theoretical values in the viscous subrange and inertial energy subrange (dashed lines). All the curves are made dimensionless by using Kolmogorov scales.}
\label{fig:enstr_struct_fun}
\end{figure}
As shown in figure \ref{fig:enstr_struct_fun}, the theoretical result for the second-order structure function of enstrophy described in equation (\ref{eq:viscous_enstr_scaling}) is well recovered by the present DNS data for small values of $r$, for all the Reynolds numbers considered in the present work. Indeed, the viscous scaling appears to constitute a good approximation for $\langle \delta \omega^2 \rangle$ only up to $r \approx 3 \eta$, while, in contrast, the profile of $\langle \delta q^2 \rangle$ starts to deviate from its theoretical viscous scaling at around $r \approx 5 \eta$, as shown in Appendix \ref{energy} (figure \ref{fig:ener_struct_fun}).

When a budget is dominated by a constant transport (e.g. of energy or enstrophy) over a certain range of scales, the relative structure function will exhibit a power law at such scales. Consistently with what is observed in the enstrophy budget (shown in figure \ref{fig:enstr_budg}), while a viscous scaling law can be observed in $\langle \delta \omega^2 \rangle$ (figure \ref{fig:enstr_struct_fun}), an inertial scaling law is not present, since the enstrophy budget is never dominated by inertial transport mechanisms, independently of the Reynolds number. On the other hand, in the inertial energy subrange, the enstrophy budget is governed by a local interscale dynamics, dominated by a scale-by-scale equilibrium between $\langle \mathcal{S} \rangle$ and $\langle \chi \rangle$, which implies a constant scaling of $\langle \delta \omega^2 \rangle$, consistently with what is visible in figure \ref{fig:enstr_struct_fun} at the large scales ($r \gg \eta$). The observed behavior is also consistent with the results reported in \cite{Simultaneous_casc_2022,Chiarini_2025}. Finally, note that the structure function $\langle \delta \omega^2 \rangle$, appears to collapse for all Reynolds numbers considered in this study, which again is consistent with the classical idea that enstrophy is a ``small-scale'' quantity and therefore exhibits universal properties. As the scale-energy (see Appendix \ref{energy}), the second-order structure function of enstrophy can be decomposed into longitudinal and transverse increments, i.e. $\delta \omega^2 = \delta \omega_{||}^2 + 2 \delta \omega_{\perp}^2$ where $\delta \omega_{||} = \delta \boldsymbol{\omega} \cdot \boldsymbol{\hat{r}}$ and $\delta \omega_{\perp} = \delta \boldsymbol{\omega} \cdot \boldsymbol{\hat{r}^{\perp}}$, respectively (the transverse increment is counted twice since it includes two directions). In statistically isotropic conditions, both the transverse increment and the structure function can be expressed as functions of the longitudinal increment as:
\begin{equation}
\begin{gathered}
\langle \delta \omega_{\perp}^2 \rangle = \frac{1}{2 r} \frac{d}{d r} \left( r^2 \langle \delta \omega_{||}^2 \rangle \right), \qquad \langle \delta \omega^2 \rangle = \frac{1}{r^2} \frac{d}{d r} \left( r^3 \langle \delta \omega_{||}^2 \rangle \right),
\label{eq:lontra_link}
\end{gathered}
\end{equation}
respectively, highlighting that the total, longitudinal and transverse structure functions are all determined by a single scalar value.

\begin{figure}[h]
\centering
\includegraphics[trim=2.5cm 0cm 2.5cm 0cm, clip, width=0.9\linewidth]{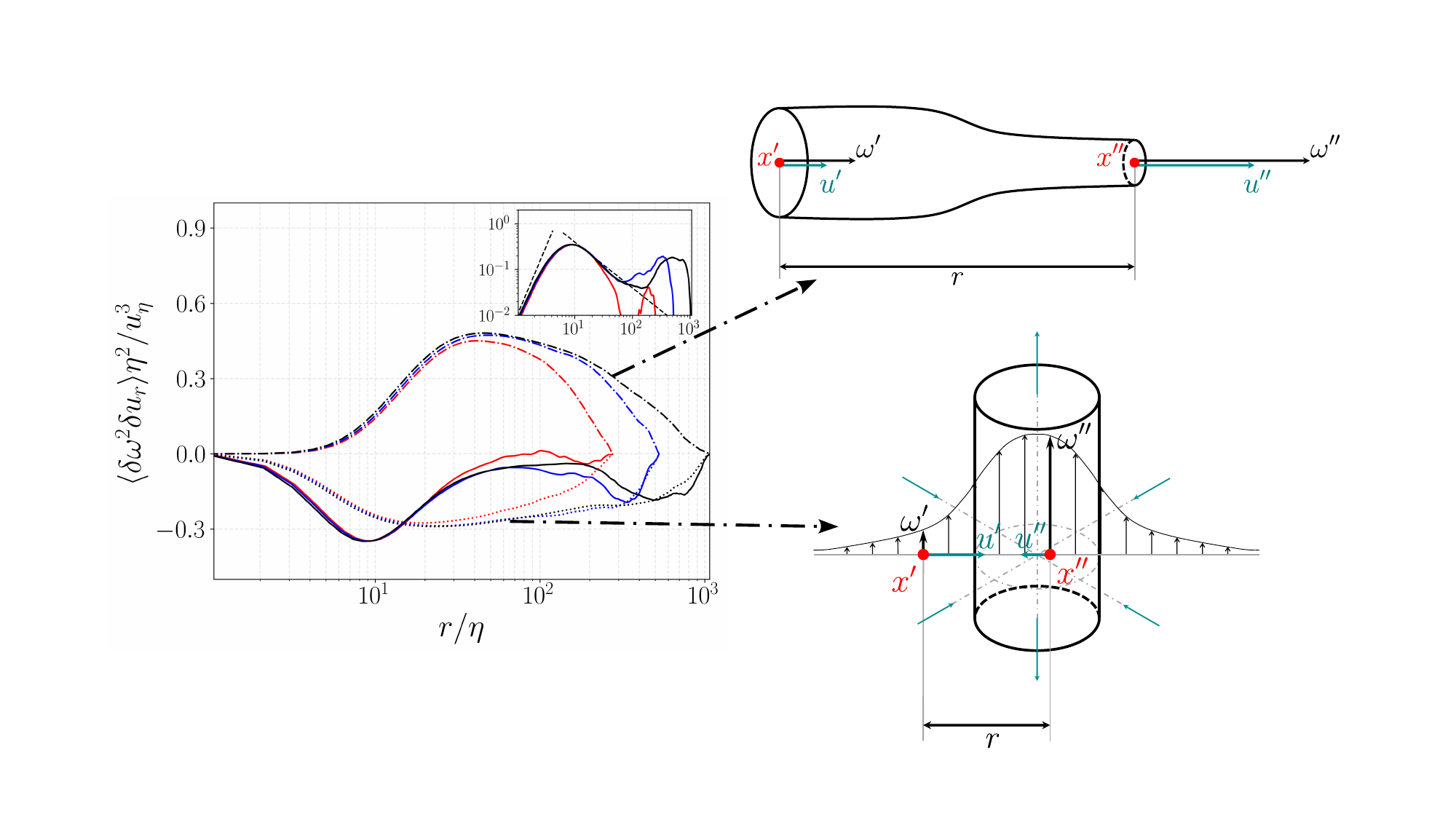}
        \put(-450,230){(a)}
        \put(-70,260){(b)}
        \put(-70,160){(c)}
        \caption{(a) Inertial enstrophy flux $\langle \delta \omega^2 \delta u_{||} \rangle$ (solid lines) with its longitudinal $\langle \delta \omega_{||}^2 \delta u_{||} \rangle$ (dash-dotted lines) and transverse $\langle \delta \omega_{\perp}^2 \delta u_{||} \rangle$ (dotted lines) contributions (main panel). The inset panel shows the inertial enstrophy flux (with the negative sign $- \langle \delta \omega^2 \delta u_{||} \rangle$) in log-log scale to highlight the viscous and inertial scalings (dashed lines), i.e., $- \langle \delta \omega^2 \delta u_{||} \rangle \sim r^{3}$ and $- \langle \delta \omega^2 \delta u_{||} \rangle = \langle \varepsilon \rangle r^{-1}$ respectively. The colors {correspond to the three different simulations, with Taylor based} Reynolds number: $Re_{\lambda} \approx 140$ (red lines), $Re_{\lambda} \approx 240$ (blue lines) and $Re_{\lambda} \approx 400$ (black lines). (b) Simplified sketch representing the longitudinal enstrophy flux scenario (see text for details). (c) Simplified sketch representing the transverse enstrophy flux scenario (see text for details).}
\label{fig:enstr_flux}
\end{figure}

By applying these relations in the diffusion-dominated range, and by exploiting isotropy, we can write the exact relation $\langle \delta \omega_{\perp}^2 \rangle = 2 \langle \delta \omega_{||}^2 \rangle$ (an analogous ratio can be derived for the velocity, see Appendix \ref{energy}). Hence, in the diffusion-dominated regime, the transverse vorticity (and velocity) increments are larger than the longitudinal ones, $\langle \delta \omega_{\perp}^2 \rangle > \langle \delta \omega_{||}^2 \rangle$. A possible physical interpretation of this is that, at these small scales, the vorticity and velocity fields remain ``coherent'' over a certain spatial extent. This coherence makes it reasonable to analyze the system using simplified schematic models. For example, considering the schematization of a vortex tube, and taking the two points $\boldsymbol{x'}$ and $\boldsymbol{x''}$ inside and aligned to the tube, we can expect a smaller increase in longitudinal vorticity than in the case where we take the two points transversely to the vortex tube (as in figure \ref{fig:enstr_flux}(b) and (c)). Analogously, for the velocity field, at small scales we may interpret $\delta u_{\perp}$ to be related to the strain rate, while $\delta u_{||}$ can be associated to a local compression or expansion event in one direction. It is noteworthy that both the ratios $\langle \delta \omega_{\perp}^2 \rangle / \langle \delta \omega_{||}^2 \rangle$ and $\langle \delta u_{\perp}^2 \rangle / \langle \delta u_{||}^2 \rangle$ are well recovered in the present data, as shown in the inset panels of figures \ref{fig:enstr_struct_fun} and \ref{fig:ener_struct_fun}.

As mentioned previously, while the energy budget is dominated by the inertial transport of $\langle \delta q^2 \rangle$ at the scales of the inertial energy subrange, the enstrophy budget at these scales simply consists of a scale-by-scale balance between production via vortex stretching and destruction via vorticity gradients. As a consequence, a power law in the velocity and a saturation in the vorticity fields are observed. In particular, as shown in the Appendix \ref{energy}, equation (\ref{eq:lontra_link_en}), for the velocity we have $\langle \delta u_{\perp}^2 \rangle / \langle \delta u_{||}^2 \rangle = 4/3$, hence implying that also in the inertial energy subrange $\langle \delta u_{\perp}^2 \rangle > \langle \delta u_{||}^2 \rangle$. However, at the scales of the inertial energy subrange, the argument concerning the small ``coherent'' motions of the viscous subrange no longer applies and the difference between longitudinal and transverse velocity increments may be given by the role of pressure. In fact, pressure can act over long distances, damping large-scale longitudinal increments and transferring this energy to transverse velocity increments. This aspect will be addressed in subsection \ref{pressure}. In contrast, at the scales associated with the inertial energy subrange, enstrophy is already saturated to the single-point value showing a plateau at a constant value of twice the rate of energy dissipation, i.e. $2 \langle \varepsilon \rangle/\nu$. This is also consistent with the scaling law for the velocity derivatives derived in \cite{nelkin1990multifractal}, $\delta \omega \sim u_{\eta} / \eta$. It is then reasonable to expect both its longitudinal and transverse contributions to be constant at these scales. Therefore, using equation (\ref{eq:lontra_link}) we can write,
\begin{equation}
\langle \delta \omega^2 \rangle = \frac{2}{\nu} \langle \varepsilon \rangle = \frac{1}{r^2} \frac{d}{d r} \left( r^3 \langle \delta \omega_{||}^2 \rangle \right),
\label{eq:rel1}
\end{equation}
leading to,
\begin{equation}
	 \langle \delta \omega_{||}^2 \rangle = \frac{2}{3 \nu} \langle \varepsilon \rangle, \qquad \langle \delta \omega_{||}^2 \rangle = \frac{\langle \delta \omega^2 \rangle}{3}.
\label{eq:rel2}
\end{equation}
Thus, the longitudinal two-point increment contributes to $1/3$ of the total scale enstrophy. Similarly, we can write,
\begin{equation}
\langle \delta \omega_{\perp}^2 \rangle = \frac{1}{2r} \frac{d}{d r} \left( r^2 \langle \delta \omega_{||}^2 \rangle \right) = \frac{1}{2r} \frac{d}{d r} \left( r^2 \frac{2}{3\nu} \langle \varepsilon \rangle \right) = \frac{2}{3\nu} \langle \varepsilon \rangle,
\label{eq:rel3}
\end{equation}
showing that the two transverse contributions account for the remaining $2/3$ of the scale enstrophy. Hence, contrary to what is observed for the velocity structure function, the longitudinal and transverse vorticity increments are equal in the inertial energy subrange, i.e.,
\begin{equation}
\langle \delta \omega_{\perp}^2 \rangle = \langle \delta \omega_{||}^2 \rangle.
\label{eq:rel4}
\end{equation}
This highlights the fact that, at the scales of the inertial energy subrange, the simple schematization based on the vortex tube used for the viscous subrange is no longer applicable. Furthermore, the vorticity equation does not include a pressure term, which therefore does not act to break the equivalence between longitudinal and transverse increments (see discussion below).

\subsection{Inertial fluxes} \label{enstr_inert_flux}
As shown in section \ref{twopoint_enstr}, the inertial enstrophy transport $\langle T_\omega \rangle$ is relatively weak compared to the other terms of the enstrophy scale-by-scale budget. Nevertheless, within the framework of equation (\ref{eq:iso_enstr}), it represents a mechanism through which interscale enstrophy transfer occurs, linking different scales of enstrophy (together with the diffusive transport). Therefore, in order to gain further insight into this interscale transfer mechanism, we analyze the inertial enstrophy flux (first term of equation (\ref{eq:iso_enstr_flux})) and its longitudinal and transverse contributions, $\langle \delta \omega^2 \delta u_r \rangle = \langle \delta \omega_{||}^2 \delta u_{||} \rangle + 2\langle \delta \omega_{\perp}^2 \delta u_{||} \rangle$, respectively (note that $\delta u_r$ and $\delta u_{||}$ are equivalent and in the following we will only refer to $\delta u_{||}$).

As shown in Appendix \ref{energy}, it is possible to derive some exact relations between the longitudinal and transverse contributions to the inertial energy flux both in the inertial and viscous subranges, namely, $\langle \delta u_{\perp}^2 \delta u_{||} \rangle / \langle \delta u_{||}^3 \rangle = 1/3$ and $\langle \delta u_{\perp}^2 \delta u_{||} \rangle / \langle \delta u_{||}^3 \rangle = 2/3$, respectively (equations (\ref{eq:tralon_ratio_energy_inert}) and (\ref{eq:tralon_ratio_energy_visc})). No similar relations exist for the enstrophy as will be show below, however it is possible to provide some estimates for the total inertial enstrophy flux. In particular, in the viscous subrange, by using the viscous scaling $\langle | \delta u_{||} | \rangle \sim \sqrt{(\langle \varepsilon \rangle / 15 \nu)}r$ and equation (\ref{eq:viscous_enstr_scaling}) we obtain:
\begin{equation}
\langle \delta \omega^2 \delta u_{||} \rangle \sim \left( \langle \chi \rangle / 3 \nu \right) \sqrt{ \langle \varepsilon \rangle / 15 \nu} \, \, r^3 \sim r^3.
\label{eq:viscous_scaling_enstr_flux}
\end{equation}
On the other hand, in the inertial energy subrange, by using the inertial scaling $\langle | \delta u| \rangle$ $\sim \langle \varepsilon \rangle^{1/3} r^{1/3}$ and $\langle | \delta \omega | \rangle \sim \langle | \delta u | \rangle / r$ $ \sim \langle \varepsilon \rangle^{1/3} r^{-2/3}$ we obtain:
\begin{equation}
\begin{split}
\langle \delta \omega^2 \delta u_{||} \rangle \sim \frac{\langle \delta u^3 \rangle}{r^2} \sim \langle \varepsilon \rangle r^{-1},
\end{split}
\label{eq:inertial_scaling_enstr_flux}
\end{equation}
similar to the relations found in \cite{Davidson2008,Simultaneous_casc_2022}.

A comment is in order. The scaling $\delta\omega\sim\delta u/r$, obtained by treating the vorticity increment as a finite-difference approximation to the velocity gradient over a separation $r$, is expected to hold only while $\omega(\mathbf{x})$ and $\omega(\mathbf{x}+\mathbf{r})$ remain correlated, i.e., for $r$ within a few tens of Kolmogorov lengths scales ($r \lesssim O(10 \eta)$), which is much smaller than the integral scale $L$ characterizing the velocity field. This is simply because vorticity is predominantly correlated within the cores of the vortical structures, which have a characteristic length scale (core diameter - $D_{ivs}$) that scales with the Kolmogorov micro-scale. Specifically, $D_{ivs} \approx 10 \eta$ \citep{JimenezWray1998,carlos_PRF2022}. Thus, for $r \gtrsim O(10 \eta)$, $\omega(\mathbf{x})$ and $\omega(\mathbf{x}+\mathbf{r})$ become statistically decorrelated and $\langle\delta\omega^2\rangle$ saturates to $\langle \delta \omega^2\rangle=2\langle\varepsilon\rangle/\nu$ irrespective of any local gradient-type scaling, whereas $\delta u(r)$ remains correlated up to separations comparable to $L$. These two behaviors thus correspond to distinct regimes and are not in conflict. The situation is different for the inertial-range enstrophy flux, $\langle\delta\omega^2\,\delta u_r\rangle$: being a triple correlation, it is not constrained to saturate to a finite value and must instead vanish as $r\to L$. Because it is not subject to the same saturation constraint, its scale dependence remains visible across the inertial range, allowing the predicted $1/r$ scaling (\ref{eq:inertial_scaling_enstr_flux}) to emerge more clearly than in the vorticity structure function itself.

Figure \ref{fig:enstr_flux} shows the inertial enstrophy flux $\langle \delta \omega^2 \delta u_{||} \rangle$ with its longitudinal $\langle \delta \omega_{||}^2 \delta u_{||} \rangle$ and transverse $\langle \delta \omega_{\perp}^2 \delta u_{||} \rangle$ contributions. Both the viscous and inertial subrange scalings described above, in equations (\ref{eq:viscous_scaling_enstr_flux}) and (\ref{eq:inertial_scaling_enstr_flux}), respectively, are found to work reasonably well (see the inset of figure \ref{fig:enstr_flux}(a)), and represent, for a given scale $r$, an overall flux of enstrophy towards smaller scales. This behaviour is consistent with the results reported in \cite{Simultaneous_casc_2022}. However, and somehow unexpectedly, the above scaling laws do not apply to either the longitudinal or the transverse contributions to the enstrophy flux, i.e., $\langle \delta \omega_{||}^2 \delta u_{||} \rangle$ and $\langle \delta \omega_{\perp}^2 \delta u_{||} \rangle$. In fact, as shown in the main panel of figure \ref{fig:enstr_flux}(a), these two terms exhibit a non trivial behavior, acting in opposite ``directions'', transporting enstrophy towards larger and smaller scales, respectively. In particular, the longitudinal enstrophy flux contribution $\langle \delta \omega_{||}^2 \delta u_{||}\rangle$ is positive, on average transporting longitudinal vorticity towards larger scales.

This behavior may be qualitatively related to the vortex stretching mechanism (schematically sketched in figure \ref{fig:enstr_flux}(b)), which both produces and transports enstrophy \citep{Davidson2008}, with vortical elements being, on average, stretched along their longitudinal direction. Indeed, as shown in figure \ref{fig:enstr_flux}(a), the longitudinal contribution to the inertial enstrophy flux, $\langle \delta \omega_{||}^2 \delta u_{||} \rangle$, peaks at intermediate scales ($r/\eta \approx 40$). Such intermediate scales allow for both a good coherence in the longitudinal vorticity, which is lost at larger scales, and the presence of significant longitudinal velocity increments, which are less significant at smaller scales. In contrast, the transverse enstrophy flux contribution, $\langle \delta \omega_{\perp}^2 \delta u_{||}\rangle$, is negative, causing an average flux from large towards small scales, which ultimately dominates the total enstrophy flux. This effect can be again qualitatively interpreted in terms of the vortex stretching mechanism which, when seen along the transverse section of the vortex element considered (as schematically depicted in figure \ref{fig:enstr_flux}(c)), is associated with a contraction, and thus a reduction in the associated enstrophy scale.

In conclusion, the vortex stretching mechanism has a dual nature acting both as an enstrophy production mechanism, through local vorticity amplification, and as an enstrophy transport mechanism, by stretching and contracting the vortical elements. In particular, the latter consists of two opposite non-trivial contributions: a longitudinal contribution causing a reverse enstrophy flux, transporting enstrophy toward larger scales, and a transverse contribution causing a direct enstrophy flux, moving enstrophy towards smaller scales. This last contribution is found to dominate the total enstrophy interscale transfer, as expected.

In the following subsection we will discuss these mechanisms in more detail, including also the role that pressure plays in such events.

\subsection{The role of pressure} \label{pressure}

In the present section, we use conditional statistics in order to better investigate the mechanisms described in the previous sections. We emphasize that the interpretation provided in this section is intended to be heuristic and is meant to provide physical insight into the mechanisms associated with intense interscale-transfer events, rather than to establish a direct causal or topological correspondence between the different quantities considered. The statistics reported are conditioned on both positive and negative ``intense'' events of longitudinal and transverse inertial energy transport $T = T_{||} + 2 T_{\perp}$ (defined in Appendix \ref{energy} through equation (\ref{eq:iso_gke})), where the selected threshold is 10 times the average. Thresholds for these terms ranging from $1$ to $1000$ have been tested, and resulted in no significant qualitative difference in the results obtained (see Appendix \ref{robustness} for further details). In the following, the conditional average operator denoted by $\langle \cdot \rangle_{10^+}$ and $\langle \cdot \rangle_{10^-}$, indicates averages taken from events with $T_{|| / \perp} > 10 |\langle T_{|| / \perp} \rangle|$ and $T_{|| / \perp} < -10 |\langle T_{|| / \perp} \rangle|$, respectively.

In order to gain further insights into the simplified scenarios depicted in figure \ref{fig:enstr_flux}(b) and (c), which represent schematic pictures of the longitudinal and transverse enstrophy transfer scenarios, we start by reporting the conditional correlations between the inertial transport of energy $T$ and enstrophy $T_{\omega}$ decomposed in their longitudinal and transverse contributions. The conditional correlations are normalized as $\langle T_{||} T_{\omega_{||}} \rangle_{10^{+/-}} /(\langle T_{||}^2 \rangle^{1/2}_{10^{+/-}} \langle T_{\omega_{||}}^2 \rangle^{1/2}_{10^{+/-}} )$ and $\langle T_{\perp} T_{\omega_{\perp}} \rangle_{10^{+/-}} /(\langle T_{\perp}^2 \rangle^{1/2}_{10^{+/-}} \langle T_{\omega_{\perp}}^2 \rangle^{1/2}_{10^{+/-}} )$. The quantities defined above are uncentered normalized cross-moments, i.e., without subtracting the corresponding means, and will hereafter be referred to simply as correlation coefficients. The moments are not centered because our interest lies in the net direction of the interscale transfer, rather than in characterizing fluctuations about the mean transfer. We have nevertheless verified that the corresponding centered normalized cross-moments exhibit the same qualitative behavior, so that the discussion that follows remains valid. The corresponding comparison is reported in Appendix \ref{robustness}.

As shown in figure \ref{fig:inert_inert_o_corr}, both the longitudinal and transverse transport events positively correlate, suggesting a link between the energy and enstrophy transport mechanisms. As expected, all the correlation coefficients are higher at small scales, where the simplified picture depicted in figures \ref{fig:enstr_flux}(b) and (c) are expected to provide a more useful representation of the underlying dynamics.

Interestingly, a positive correlation is observed between the longitudinal inertial transports of energy and enstrophy, even though their average values have opposite signs ($\langle T_{||} \rangle > 0$ and $\langle T_{\omega_{||}} \rangle < 0$), in agreement with the behavior of the longitudinal inertial fluxes reported in figures \ref{fig:enstr_flux}(a) and \ref{fig:ener_flux}. The reason for this is that on average, for events of negative longitudinal energy transport (i.e. reverse transport), the enstrophy transport is much more intense than in cases of positive (i.e. direct) energy transport (not shown here). This is consistent with what is observed in the literature \cite{Ashurst_1987,Tsinober_Kit_Dracos_1992,tsinober2009informal,buaria2020vortex,Johnson_annurev}.

\begin{figure}[ht!]
\centering
\includegraphics[width=0.6\linewidth]{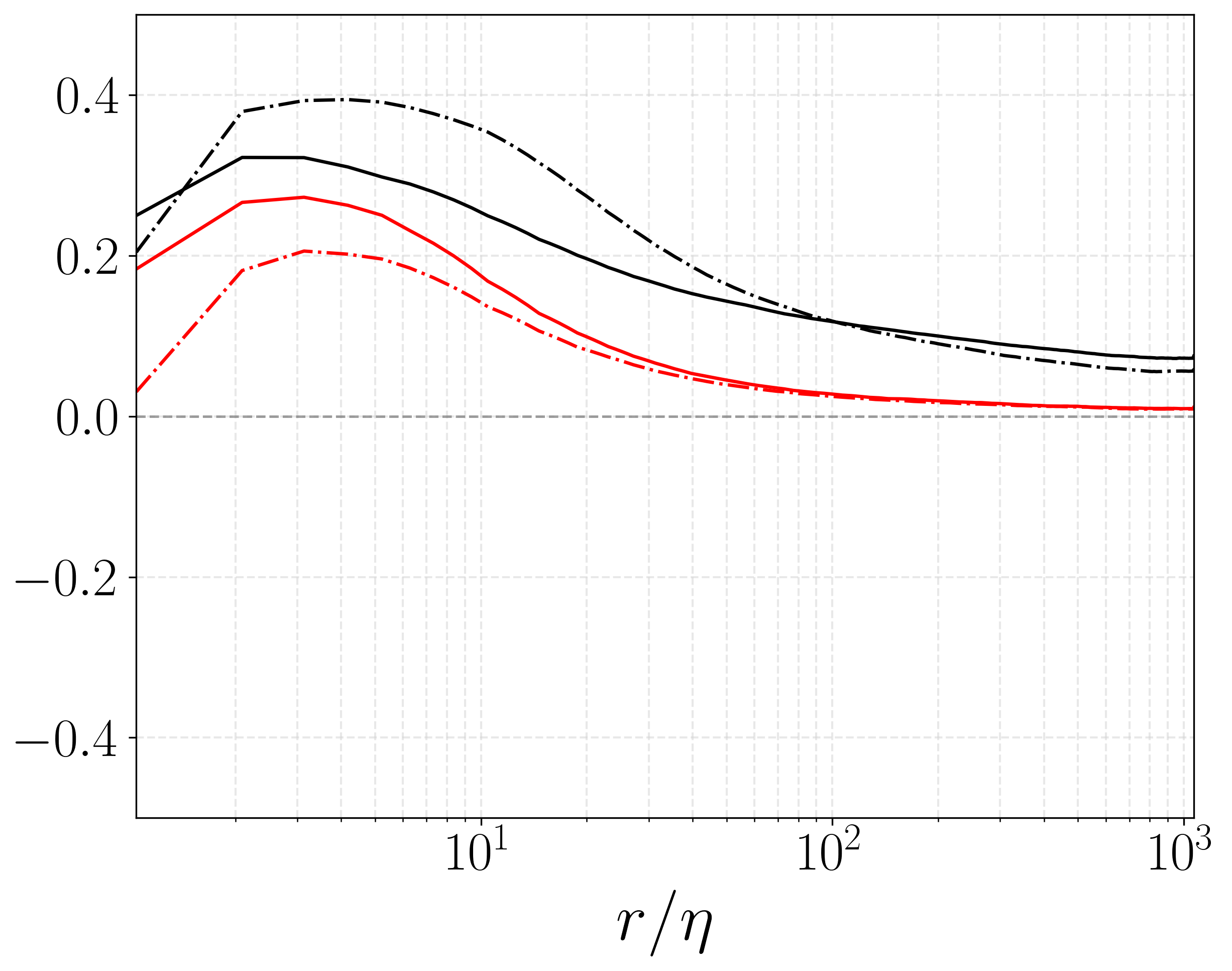}
        \caption{ Conditional correlation coefficients between longitudinal energy and enstrophy transports $\langle T_{||} T_{\omega_{||}} \rangle_{10^{+/-}} /(\langle T_{||}^2 \rangle^{1/2}_{10^{+/-}} \langle T_{\omega_{||}}^2 \rangle^{1/2}_{10^{+/-}} )$ (black lines) and transverse energy and enstrophy transports $\langle T_{\perp} T_{\omega_{\perp}} \rangle_{10^{+/-}} /(\langle T_{\perp}^2 \rangle^{1/2}_{10^{+/-}} \langle T_{\omega_{\perp}}^2 \rangle^{1/2}_{10^{+/-}} )$ (red lines), conditioned on intense positive events of energy transport $\langle \cdot \rangle_{10^+}$ (solid lines), i.e. direct energy transport, and intense negative events of energy transport $\langle \cdot \rangle_{10^-}$ (dash-dotted lines), i.e. reverse energy transport.}
\label{fig:inert_inert_o_corr}
\end{figure}

Further insight into the events governing energy and enstrophy transport, previously shown to be positively correlated, can be obtained by examining the role of pressure in these processes~\cite{buaria2023role}. In fact, the pressure transport term in the energy transport equation, can be used as a proxy to the transport terms analyzed in the pervious sections since it contains information about the dynamics of the events considered. In the inhomogeneous evolution equation for the velocity structure function $\langle \delta q^2 \rangle$ \cite{Hill2002exact}, the pressure transport term takes the form:
\begin{equation}
\langle T_p \rangle = - \frac{2}{\rho} \frac{\partial \langle \delta p \delta u_i \rangle}{\partial x_{c_i}},
\end{equation}
thus, resulting in an average energy transport only in the physical space, which is zero in statistically homogeneous conditions. However, the pressure transport term can be rewritten as:
\begin{equation}
\begin{gathered}
T_p = - \frac{2}{\rho} \frac{\partial \delta p \delta u_i}{\partial x_{c_i}} = - \frac{2}{\rho} \delta u_i \,\delta \! \left( \frac{\partial p}{\partial x_i} \right) = \underbrace{- \frac{2}{\rho} \delta u_{||} \, \delta \! \left( \frac{\partial p}{\partial x_{||}} \right)}_{T_{p ||}} \underbrace{- \frac{4}{\rho} \delta u_{\perp} \, \delta \! \left( \frac{\partial p}{\partial x_{\perp}} \right)}_{2 T_{p \perp}}.
\end{gathered}
\label{eq:p_transp}
\end{equation}
In particular, the longitudinal contribution $T_{p ||}$ may help us in interpreting the kinematics of some archetypal events that are relevant for both the energy and the enstrophy transport.

\begin{figure}[ht!]
\centering
\includegraphics[trim=4cm 0cm 4.5cm 0cm, clip, width=0.9\linewidth]{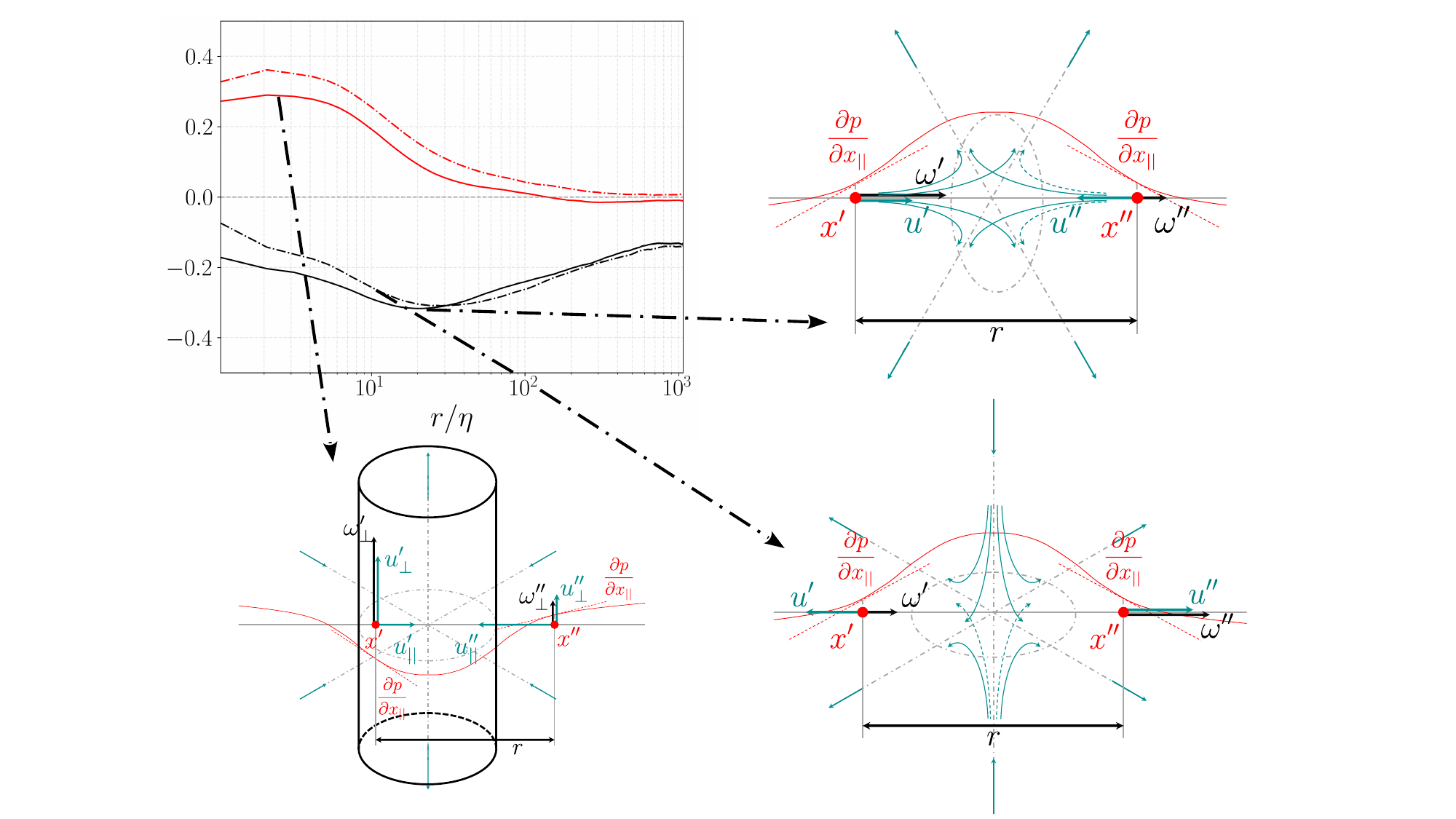}
        \put(-475,315){(a)}
        \put(-35,315){(b)}
        \put(-440,140){(c)}
        \put(-35,140){(d)}
        \caption{ (a) Conditional correlation coefficients between the longitudinal pressure transport and the longitudinal energy transport contribution, $\langle T_{||} T_{p ||} \rangle_{10^{+/-}} /(\langle T_{||}^2 \rangle^{1/2}_{10^{+/-}} \langle T_{p ||}^2 \rangle^{1/2}_{10^{+/-}} )$ (black lines), and between the longitudinal pressure transport and the transverse energy transport contribution, $\langle T_{\perp} T_{p ||} \rangle_{10^{+/-}} /(\langle T_{\perp}^2 \rangle^{1/2}_{10^{+/-}} \langle T_{p ||}^2 \rangle^{1/2}_{10^{+/-}} )$ (red lines). Both correlation coefficients are conditioned on intense positive energy transport events (direct transfer), $\langle \cdot \rangle_{10^+}$ (solid lines), and on intense negative energy transport events (reverse transfer), $\langle \cdot \rangle_{10^-}$ (dash-dotted lines). (b) to (d): Schematic sketches representing a direct longitudinal energy transport (b), direct transverse energy transport (c), and a reverse longitudinal energy transport (d).}
\label{fig:inert_p_corr}
\end{figure}

Figure \ref{fig:inert_p_corr}(a) shows the correlation coefficients between the longitudinal pressure transport $T_{p ||}$ and the longitudinal and transverse energy transport contributions during intense events of direct and reverse energy transport ($\langle T_{||} T_{p ||} \rangle_{10^{+/-}}$ and $\langle T_{\perp} T_{p ||} \rangle_{10^{+/-}}$). Intense longitudinal energy transport events, both positive and negative, are negatively correlated with longitudinal pressure transport ($\langle T_{||} T_{p ||}\rangle_{10^{+/-}} < 0$), as shown by the black lines in figure \ref{fig:inert_p_corr}(a). For a positive longitudinal energy transport event ($T_{||} > 0$, corresponding to direct transport), this implies a negative longitudinal pressure transport ($T_{p ||} < 0$). Since $T_{p ||} = -(2 / \rho) \delta u_{||} \delta (\partial p / \partial x_{||})$, a negative value of $T_{p ||}$ requires $\delta u_{||}$ and $\delta (\partial p / \partial x_{||})$ to have the same sign. In direct energy transport events, $\delta u_{||}$ is expected to be negative (see figure \ref{fig:inert_p_corr}(b)), which in turn implies $\delta (\partial p / \partial x_{||}) < 0$. In the simplified picture illustrated in figure \ref{fig:inert_p_corr}(b), this sign of the pressure-gradient increment is consistent with the presence of a high-pressure region between the two points $\boldsymbol{x'}$ and $\boldsymbol{x''}$ (see again figure \ref{fig:inert_p_corr}(b)).
The emerging scenario is consistent with a simplified picture of an inertial event involving a strong compression in the longitudinal direction causing a high pressure zone and redistributing the longitudinal energy component into the transverse directions. This is analogous to $\lambda_2 > 0$ events, that are related to strain self-amplification mechanisms \citep{annurevMeneveau2011,Johnson_annurev}.

On the other hand, for negative longitudinal energy transport events ($T_{||} < 0$, corresponding to inverse transport; see figure \ref{fig:inert_p_corr}(d)), $\delta u_{||}$ is expected to be positive. In this case, the negative correlation implies a positive pressure transport ($T_{p ||} > 0$). Since $T_{p ||} = -(2 / \rho) \delta u_{||} \delta (\partial p / \partial x_{||})$, a positive value of $T_{p ||}$ with $\delta u_{||} > 0$ requires $\delta (\partial p / \partial x_{||}) < 0$. Within the simplified picture illustrated in figure \ref{fig:inert_p_corr}(d), this sign of the pressure-gradient increment is again consistent with a higher-pressure region between the two points $\boldsymbol{x'}$ and $\boldsymbol{x''}$. The observed behavior may likewise be associated with $\lambda_2 > 0$ events, with a different spatial arrangement. In this second schematic picture, the direction of the single intense compression is aligned along one of the transverse directions, while the two points $\boldsymbol{x'}$ and $\boldsymbol{x''}$ lie along one of the directions of expansion, as illustrated in figure \ref{fig:inert_p_corr}(d).

The idea that these events are the most relevant for longitudinal energy transport is also consistent with the negative mean value of $\delta u_{||}^3$, indicating a net direct longitudinal energy transfer. Indeed, in the schematic sketches illustrated in figure \ref{fig:inert_p_corr}(b) and (d), compressions ($\delta u_{||} < 0$) are smaller in number but greater in intensity than expansions ($\delta u_{||} > 0$), whose intensity is distributed over two directions. As a result, this leads to a negative skewness $\langle \delta u_{||}^3 \rangle < 0$. As shown in figure \ref{fig:inert_p_corr}(a), this anti-correlation between longitudinal energy and pressure transports also holds at large scales, as expected given the capability of pressure to act non-locally.

While longitudinal energy transport events are negatively correlated with pressure transport, intense transverse energy transport events (both positive and negative) show a positive correlation with longitudinal pressure transport at small scales ($\langle T_{\perp} T_{p ||} \rangle_{10^{+/-}} > 0$), as indicated by the red lines in figure \ref{fig:inert_p_corr}(a). Following heuristic arguments analogous to those presented above, this positive correlation suggests $\delta (\partial p / \partial x_{||}) > 0$. Within the schematic sketch illustrating a simplified positive energy transport scenario in figure \ref{fig:inert_p_corr}(c), this may correspond to the presence of a low-pressure region between the two points $\boldsymbol{x'}$ and $\boldsymbol{x''}$. This low-pressure region may be associated with the rotation of a vortical element (also illustrated in figure \ref{fig:inert_p_corr}(c)), a picture further supported by the positive correlation, at comparable scales, between transverse energy and enstrophy transports in intense energy transport events $\langle T_{\perp} T_{\omega \perp} \rangle_{10^{+/-}} > 0$ (figure \ref{fig:inert_inert_o_corr}). Together, these results suggest that the vortex stretching mechanism may be particularly relevant in transverse energy transport. Hence, transverse energy transport seems to be more effectively accomplished by small-scale processes such as vortex stretching mechanisms (as in figure \ref{fig:inert_p_corr}(c)), rather than by large-scale inertial compressions, which are instead more relevant in longitudinal energy transfer.

This interpretation is consistent with the scale-dependent structure of the inertial energy flux and of its longitudinal and transverse contributions $\langle \delta q^2 \delta u_{||} \rangle = \langle \delta u_{||}^3 \rangle + 2 \langle \delta u_{\perp}^2 \delta u_{||} \rangle$, with the factor of 2 accounting for the two transverse directions. In fact, in the inertial subrange, the flux is dominated by the longitudinal contribution, $\langle \delta u_{||}^3 \rangle > 2 \langle \delta u_{\perp}^2 \delta u_{||} \rangle$ (see Appendix \ref{energy}), reflecting the stronger role played by large-scale inertial compressions in the longitudinal energy transfer. Conversely, in the viscous subrange the transverse contribution becomes dominant, $2 \langle \delta u_{\perp}^2 \delta u_{||} \rangle > \langle \delta u_{||}^3 \rangle$ (see again Appendix \ref{energy}), indicating the relevance of small-scale mechanisms involved in the transverse energy transfer.

\section{Conclusions} \label{concl}

In the present work, we analyze two-point enstrophy $\delta \omega^2 = \delta \omega_i \delta \omega_i$ in stationary homogeneous isotropic turbulence at different Reynolds numbers, in the range $140 \lesssim Re_{\lambda} \lesssim 400$. The investigation shows that, at scales $r > 10 \eta$, the enstrophy budget is entirely determined by production via vortex stretching, which balances enstrophy destruction via vorticity gradients. Unlike the energy budget, the enstrophy budget is found to never be dominated by the inertial transport, since the vortex stretching production term is active up to the diffusion-dominated scales, independently of the Reynolds number. Consequently, the second-order structure functions of vorticity exhibit a plateau in the inertial energy subrange, implying also $\langle \delta \omega^2 \rangle$ to be equally distributed among its longitudinal and transverse contributions $\langle \delta \omega_{||}^2 \rangle = \langle \delta \omega_{\perp}^2 \rangle$ (with $\langle \delta \omega^2 \rangle = \langle \delta \omega_{||}^2 \rangle + 2 \langle \delta \omega_{\perp}^2 \rangle$). This is in contrast to the viscous subrange, where the transverse enstrophy contribution is dominant $\langle \delta \omega_{\perp}^2 \rangle = 2 \langle \delta \omega_{||}^2 \rangle$.

Despite its small relevance in the overall behavior of the enstrophy budget, the inertial enstrophy flux $\langle \delta \omega^2 \delta u_{||} \rangle$ provides the main mechanism that links enstrophy across different scales alongside diffusive transport. The study of the inertial enstrophy flux and its longitudinal and transverse contributions $\langle \delta \omega^2 \delta u_{||} \rangle = \langle \delta \omega_{||}^2 \delta u_{||} \rangle + 2 \langle \delta \omega_{\perp}^2 \delta u_{||} \rangle$ provided a deeper insight in the enstrophy dynamics, highlighting the dual nature of the vortex stretching mechanism, which, in addition to producing enstrophy by locally amplifying vorticity, also acts as a transport mechanism. The net effect is a flux of enstrophy from large to small scales, with scaling laws of $\langle \delta \omega^2 \delta u_{||} \rangle \sim r^3$ in the viscous subrange and $\langle \delta \omega^2 \delta u_{||} \rangle \sim \langle \varepsilon \rangle r^{-1}$ in the inertial energy subrange. However, this overall direct enstrophy flux from large to small scales, is found to be composed of a non-trivial combination of longitudinal and transverse contributions. In fact, the longitudinal enstrophy flux $\langle \delta \omega_{||}^2 \delta u_{||} \rangle$ provides a reverse contribution, transferring enstrophy from small to large scales through the average stretching of vortical elements in the longitudinal direction. In contrast, the transverse enstrophy flux $\langle \delta \omega_{\perp}^2 \delta u_{||} \rangle$, is found to give a direct contribution, transferring enstrophy from large to small scales through the lateral contraction of vortical elements in vortex stretching events, on average.

As suggested by the conditional correlations, the phenomenology of the longitudinal and transverse transports of energy and enstrophy seems to be analogous, at least at the small ``coherent'' scales. In particular, transverse energy transport events seem to be linked to local low-pressure regions that may be attributed to the presence of vortical elements. This may suggest that the vortex stretching mechanism is particularly relevant for the transverse energy transport. On the other hand, longitudinal energy transport events are found to correlate with local high-pressure regions, suggesting that the most relevant events in longitudinal energy transport are large-scale inertial compressions. In such events, the pressure redistributes energy from the longitudinal contribution $\langle \delta u_{||}^2 \rangle$ to the transverse one $\langle \delta u_{\perp}^2 \rangle$, thus allowing a phenomenological interpretation for the uneven distribution of energy between its longitudinal and transverse contributions $\langle \delta u_{\perp}^2 \rangle > \langle \delta u_{||}^2 \rangle$.

\begin{acknowledgments}
The authors would like to thank Prof. Guido Boffetta and Prof. Stefano Musacchio for their helpful comments and valuable exchange of ideas. We also thank Dr. Alessandro Chiarini for the insightful discussions on the topic. This work has been supported by the Italian Ministry of University and Research (MUR), through the PRIN 2022 PNRR Project 20224NY2YK - ADMIRE, CUP E53C24002710006, funded by the National Recovery and Resilience Plan (PNRR), Italy, Mission 4 Component C2 Investment 1.1.
\end{acknowledgments}

\appendix
\section{Direct Numerical Simulations} \label{dataset}
The direct numerical simulations (DNS) on which the analysis is carried out are the same as those presented in \cite{carlos_PRF2022} (see Table \ref{table_param}). In particular, the present analysis uses three simulations of statistically stationary (forced) homogeneous isotropic turbulence (HIT) carried out in a triply-periodic domain with sizes $L = 2 \pi$ using classical pseudo-spectral methods, and with a Taylor-based Reynolds number between $140 \lesssim Re_{\lambda} \lesssim 400$. Statistical stationarity is obtained by the imposition of an external power input of the forcing that (on average) balances the viscous dissipation of kinetic energy, using the method described in \cite{Alvelius99}. The simulations use a number of collocation points between $512^3 \lesssim N^3 \lesssim 2048^3$, while the resulting resolution is $k_{max} \eta \approx 2.0$ for all simulations. Further details are given in \cite{carlos_PRF2022}.

\begin{table} [ht!]
        \begin{center}
                \def~{\hphantom{0}}
                \begin{tabular}{ccccccccc}
$N^3$ & $\nu$ & $Re_{\lambda}$ & $K$ & $L_{11}$ & $\varepsilon$ & $\lambda$ & $\eta$  & $k_{max}\eta$ \\
\hline
$512^3$ & $0.0057$     & $136$ & $13.31$ &  $0.83$ & $11.18$ & $0.260$   & $0.0113$ & $ 1.94$ \\
$1024^3$ & $0.0023$   & $241$ & $13.96$ &  $0.84$ & $9.74$   & $0.182$   & $0.00594$ & $ 2.03$ \\
$2048^3$ & $0.0010$ & $393$ & $17.68$ &  $0.96$ & $13.47$   & $0.115$   & $0.00294$ & $ 2.00$ \\
                \end{tabular}
                \caption{
                Physical and computational parameters of the DNS of forced (statistically stationary) HIT carried out in this work (the reported statistics correspond to the final instant of each simulation). Total number of collocation points ($N^3$); Kinematic viscosity ($\nu$); Taylor based Reynolds number ($Re_{\lambda}$); Turbulent kinetic energy ($K$); Longitudinal integral length scale ($L_{11}$); Viscous dissipation rate ($\varepsilon$); Kolmogorov microscale ($\eta$); Taylor-microscale ($\lambda$); Resolution ($k_{max}\eta$).
                }
                \label{table_param}
        \end{center}
\end{table}

\section{Scale-energy transport equation and scaling} \label{energy}

\begin{figure}[ht!]
\centering
\includegraphics[width=0.6\linewidth]{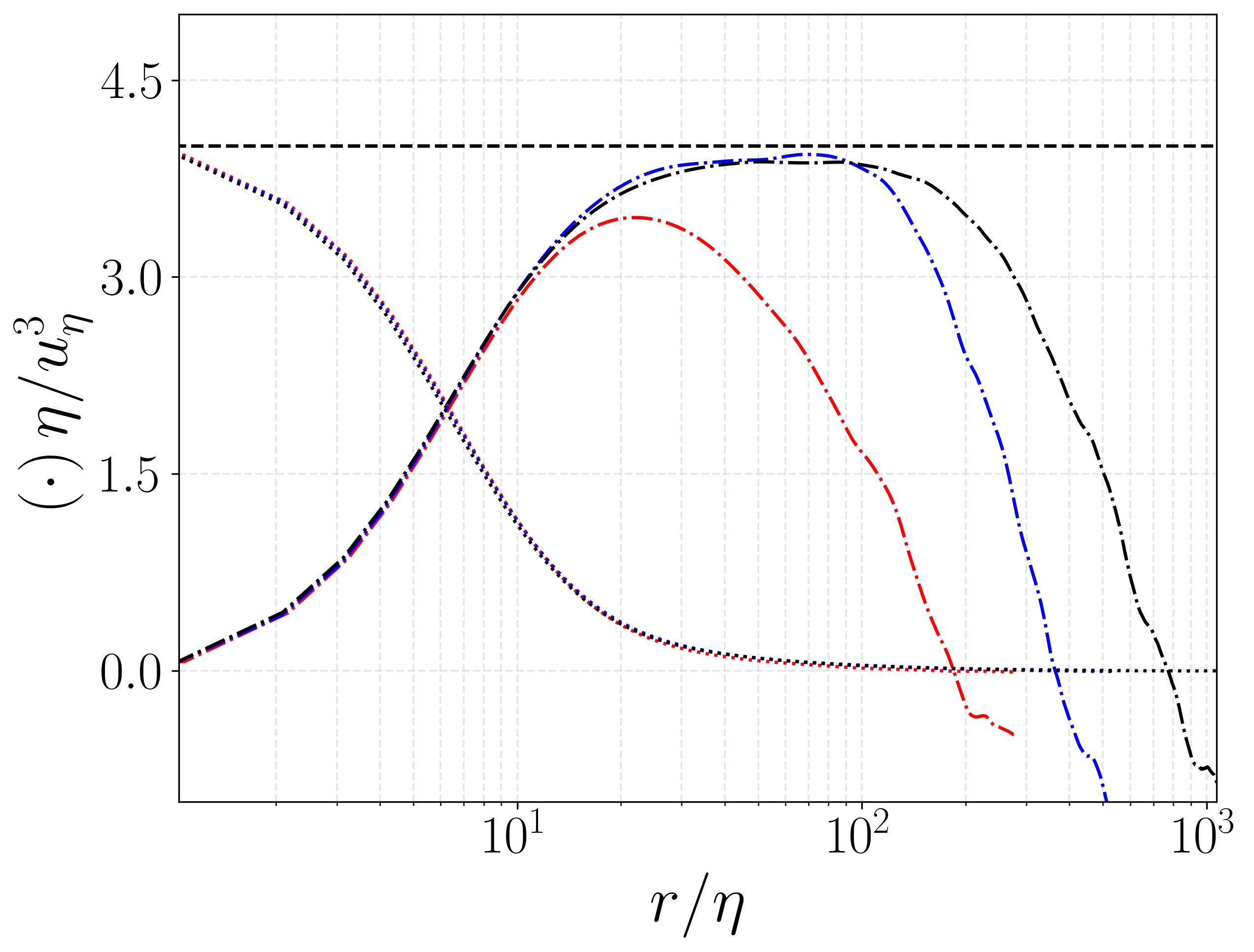}
        \caption{ Budget terms of $\langle \delta q^2\rangle$ (equation (\ref{eq:iso_gke})): viscous dissipation $4 \langle \varepsilon \rangle$ (dashed lines), diffusive transport $\langle D \rangle$ (dotted lines) and inertial transport $\langle T \rangle$ (dash-dotted lines). The colors indicate the Taylor-scale Reynolds number: $Re_{\lambda} \approx 140$ (red lines), $Re_{\lambda} \approx 240$ (blue lines) and $Re_{\lambda} \approx 400$ (black lines). All the curves are made dimensionless by using Kolmogorov units.}
\label{fig:energy_budg}
\end{figure}

In this appendix, we report an analysis of the second-order structure function of the velocity field $\langle \delta q^2 \rangle = \langle \delta u_i \delta u_i \rangle$, aimed at providing a comparison with respect to two-point enstrophy $\langle \delta \omega^2 \rangle$ analyzed in the main body of the text. The formalism used here is the one offered by the generalized Kolmogorov equation (GKE), which governs the evolution of the second-order structure function of the velocity $\delta q^2$. In statistically stationary, homogeneous and isotropic conditions this equation reads:
\begin{equation}
\begin{gathered}
-\underbrace{\frac{1}{r^2}\frac{d}{d r} \bigg ( r^2 \langle \delta q^2 \delta u_r \rangle \bigg )}_{\langle T \rangle} + \underbrace{2\nu \frac{1}{r^2} \frac{d}{d r} \left ( r^2 \frac{d \langle \delta q^2 \rangle}{d r} \right )}_{\langle D \rangle} = 4 \langle \varepsilon \rangle + 2 \langle \delta u_i \delta f_i \rangle
\end{gathered}
\label{eq:iso_gke}
\end{equation}
In equation (\ref{eq:iso_gke}), $T$ and $D$ are the inertial and diffusive energy transports in the space of scales, while $\varepsilon$ is the viscous dissipation. The last term represents the injection of energy imposed by the large-scale forcing. The divergence terms (on the LHS) are expressed in spherical coordinates $\boldsymbol{r} = (r, \phi, \theta)$ in order to highlight the dependence of the scale-space fluxes on the sole radial direction $r$ in homogeneous isotropic conditions, while the source/sinks terms (on the RHS) are written in Cartesian coordinates for the sake of simplicity. $T$ and $D$ are inertial and diffusive energy transports in the scale space and $\varepsilon$ is the dissipation. In particular, the inertial energy transport can be split into longitudinal and transverse contributions $T = T_{||} + 2 T_{\perp}$, respectively, where $T_{||} = - (1/ r^2) d / dr (r^2 \delta u_{||}^3)$ and $T_{\perp} = - (1/ r^2) d / dr (r^2 \delta u_{\perp}^2 \delta u_{||} )$. The terms of equation (\ref{eq:iso_gke}) are shown in figure \ref{fig:energy_budg} for the three different Reynolds numbers considered in this study.

As is well known, it is possible to identify a range of scales over which the energy budget is dominated by the diffusive transport (the viscous subrange) and, for sufficiently high Reynolds numbers, one over which the budget is dominated by the inertial transport (the inertial energy subrange). Accordingly, and as shown in figure \ref{fig:energy_budg}, while the diffusive transport terms collapse for all the Reynolds numbers (being dominant at small-scales), the inertial transport shows a wider and wider range of activity as the Reynolds number increases, connecting the regions of energy injection (at large scales) to the regions where diffusive phenomena are relevant(at the small scales). Thus, the classical widening of the inertial energy subrange with the Reynolds number is observed.

By integrating equation (\ref{eq:iso_gke}) over a spherical volume in the space of scales ($r$), and by considering scales small enough not to be affected by the external power input ($r \ll L_{11}$), the two-point energy budget can be written in terms of fluxes as,
\begin{equation}
\langle \delta q^2 \delta u_r \rangle - 2\nu \frac{d \langle \delta q^2 \rangle}{d r}  = - \frac{4}{3} \langle \varepsilon \rangle r,
\label{eq:iso_flux_gke}
\end{equation}
thus showing that the essential feature of turbulence in statistically stationary and homogeneous isotropic conditions is the single process of scale-energy flux from large to small scales. Notice that this energy flux is radial in the space of scales, linear with the separation $r$ and proportional to the viscous dissipation rate $\langle \varepsilon \rangle$.

By considering only scales within the viscous subrange ($r / \eta \approx 1$) where $\langle \delta q^2 \delta u_r \rangle \approx 0$, and by integrating equation (\ref{eq:iso_flux_gke}) on $r$, it is possible to obtain the viscous scaling law for the second-order velocity structure function:
\begin{equation}
\langle \delta q^2 \rangle = \frac{1}{3 \nu} \langle \varepsilon \rangle r^2.
\label{eq:viscous_scaling}
\end{equation}

\begin{figure}[ht!]
\centering
\includegraphics[width=0.6\linewidth]{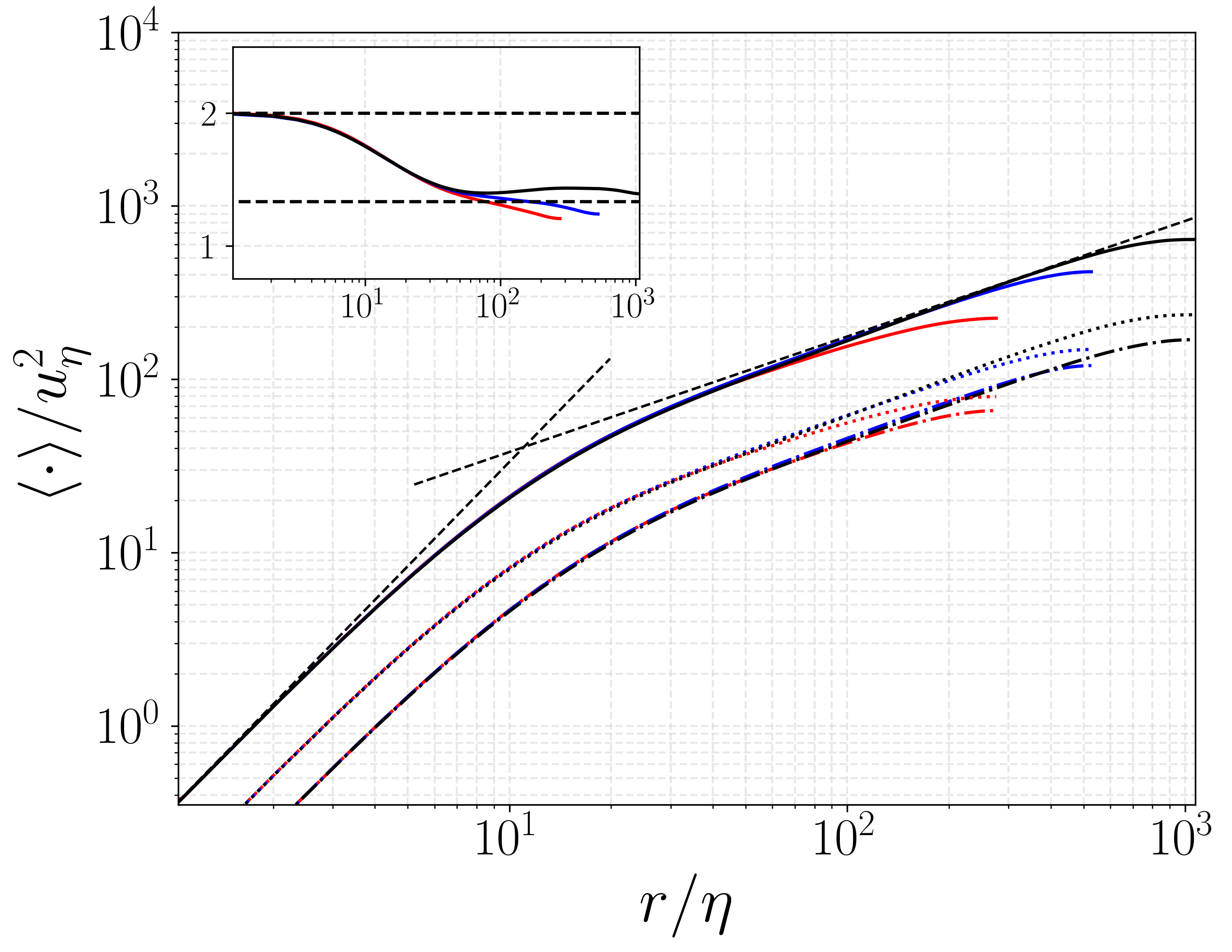}
	\caption{ Second-order structure function of energy, $\langle \delta q^2 \rangle$ (solid lines), longitudinal contribution $\langle \delta u_{||}^2 \rangle$ (dash-dotted lines) and transverse contribution $\langle \delta u_{\perp}^2 \rangle$ (dotted lines). The viscous and the inertial scaling laws $\langle \delta q^2 \rangle = 1/(3 \nu) \langle \varepsilon \rangle r^2$ and $\langle \delta q^2 \rangle = C' \langle \varepsilon \rangle^{2/3} r^{2/3}$ are reported (dashed black lines). The colors corresponds to the three different simulations, with Taylor based Reynolds numbers of $Re_{\lambda} \approx 140$ (red lines), $Re_{\lambda} \approx 240$ (blue lines) and $Re_{\lambda} \approx 400$ (black lines). The inset panel shows the ratio $\langle \delta u_{\perp}^2 \rangle / \langle \delta u_{||}^2 \rangle$ for the three simulations, together with the theoretical values in the viscous and inertial subranges (dashed lines). All the curves are made dimensionless by using Kolmogorov scales.}
\label{fig:ener_struct_fun}
\end{figure}

Figure \ref{fig:ener_struct_fun} shows the second-order velocity structure function, $\langle \delta q^2 \rangle$, for all the DNS considered in the present work. As expected, the above scaling law is well recovered for all the Reynolds numbers considered, as shown by the good collapse obtained between all the curves at small scales.

Similarly, since the second-order velocity structure function can be decomposed into a longitudinal and a transverse contribution, $\langle \delta q^2 \rangle = \langle \delta u_{||}^2 \rangle + 2 \langle \delta u_{\perp}^2 \rangle$, respectively, equation (\ref{eq:iso_gke}) can be used to write (again for scales $r \ll L_{11}$), a governing equation for the longitudinal velocity increments,
\begin{equation}
\begin{gathered}
        \frac{1}{3r^3 } \frac{d}{d r} \bigg ( r^4 \langle \delta u_{||}^3 \rangle \bigg ) - 2\nu \frac{d}{d r} \bigg [ \frac{1}{r^2} \frac{d}{d r} \bigg ( r^3 \langle \delta u_{||}^2 \rangle \bigg ) \bigg ] = - \frac{4}{3} \langle \varepsilon \rangle r,
\label{eq:iso_lon_gke}
\end{gathered}
\end{equation}
where each term has the same physical meaning as in equation (\ref{eq:iso_flux_gke}) (with the forcing term discarded since $r \ll L_{11}$).

Analogously, by limiting equation (\ref{eq:iso_lon_gke}) to the small scales of motion ($r \approx \eta$) and by integrating in $r$ twice, it is possible to obtain the well-known viscous scaling law for the second-order longitudinal velocity structure function:
\begin{equation}
        \langle \delta u_{||}^2 \rangle = \frac{1}{15 \nu} \langle \varepsilon \rangle r^2.
\label{eq:visc_lon_scal}
\end{equation}

By exploiting this scaling law, together with the following isotropic relations:
\begin{equation}
\begin{gathered}
\langle \delta u_{\perp}^2 \rangle = \frac{1}{2 r} \frac{d}{d r} \left( r^2 \langle \delta u_{||}^2 \rangle \right), \qquad \langle \delta q^2 \rangle = \frac{1}{r^2} \frac{d}{d r} \left( r^3 \langle \delta u_{||}^2 \rangle \right),
\label{eq:lontra_link_en}
\end{gathered}
\end{equation}
it is possible to obtain the following relation between the longitudinal and transverse velocity structure functions, within the viscous subrange:
\begin{equation}
\langle \delta u_{\perp}^2 \rangle = 2 \langle \delta u_{||}^2 \rangle.
\label{eq:visc_en_ratio}
\end{equation}

\begin{figure}[t!]
\centering
\includegraphics[width=0.6\linewidth]{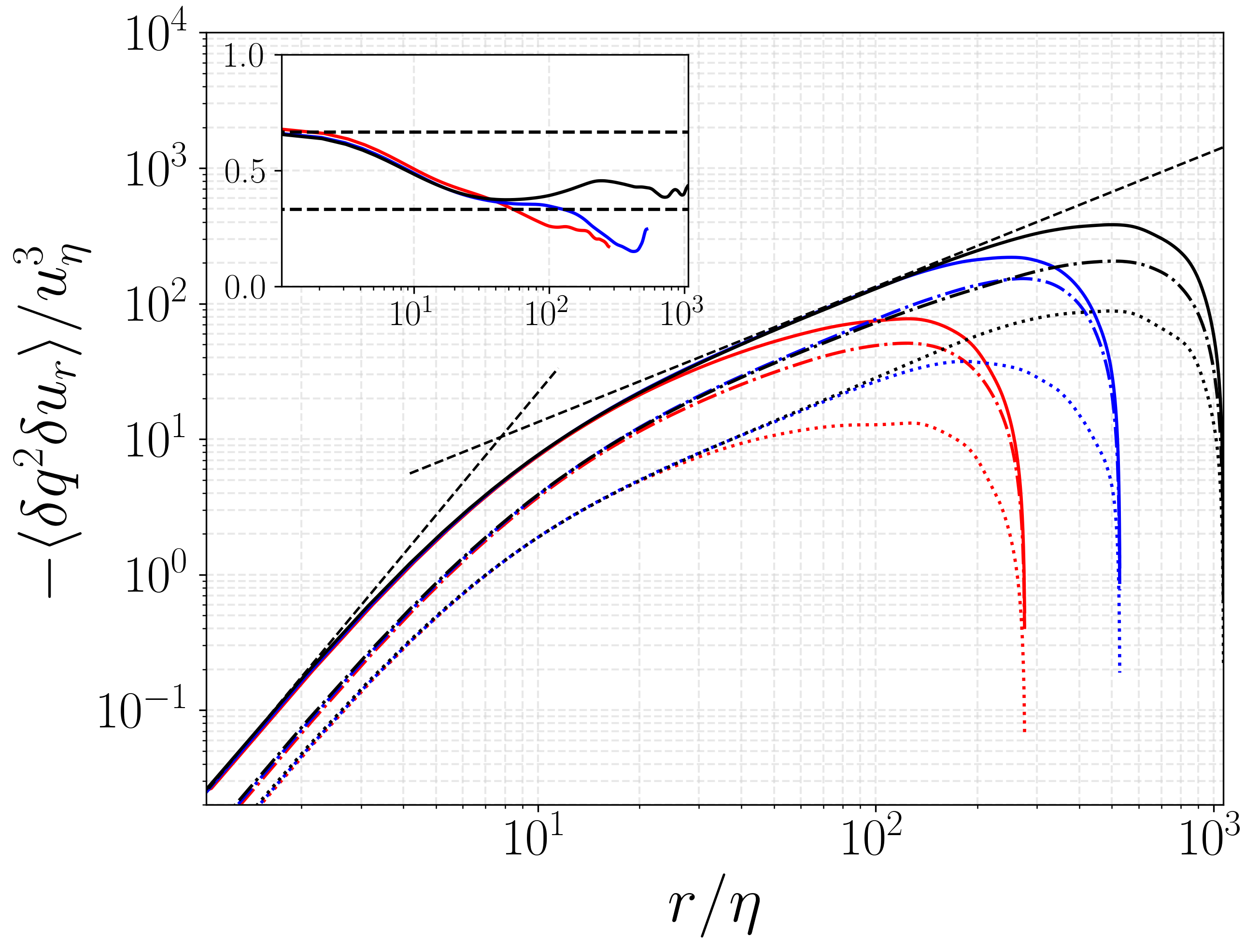}
	\caption{ Inertial energy flux, $- \langle \delta q^2 \delta u_{||} \rangle$ (solid lines), longitudinal contribution $-\langle \delta u_{||}^3 \rangle$ (dash-dotted lines) and transverse contribution $-\langle \delta u_{\perp}^2 \delta u_{||} \rangle$ (dotted lines). The viscous and the inertial scaling laws $- \langle \delta q^2 \delta u_{||} \rangle \sim r^3$ and $- \langle \delta q^2 \delta u_{||} \rangle = 4/3 \langle \varepsilon \rangle r$ are reported as dashed black lines. The colors correspond to the three different simulations, with Taylor based Reynolds numbers of $Re_{\lambda} \approx 140$ (red lines), $Re_{\lambda} \approx 240$ (blue lines) and $Re_{\lambda} \approx 400$ (black lines). In the inset panel, the ratio $\langle \delta u_{\perp}^2 \delta u_{||} \rangle / \langle \delta u_{||}^3 \rangle$ is reported together with its theoretical values in the viscous and inertial subranges (dashed lines). All the curves are made dimensionless by using Kolmogorov scales.}
    \label{fig:ener_flux}
\end{figure}

On the other hand, for sufficiently high Reynolds numbers, a range emerges such that $\eta \ll r \ll L$. This range is called the ``inertial energy subrange'' and is composed of scales that are far from both the energy injection scales and those at which viscous effects are non-negligible. In this inertial subrange, inviscid transfer of energy dominates the flow dynamics. By considering equation (\ref{eq:iso_lon_gke}) in the inertial subrange, and by integrating it in $r$, one obtains the famous \textit{four-fifth} scaling law:
\begin{equation}
\langle \delta u_{||}^3 \rangle = - \frac{4}{5} \langle \varepsilon \rangle r.
\label{eq:four_fifth}
\end{equation}

Generalization of this law to second-order longitudinal structure function leads to the famous \textit{two-third} scaling law \citep{kolmogorov41b}:
\begin{equation}
\langle \delta u_{||}^2 \rangle \sim \langle \varepsilon \rangle^{2/3} r^{2/3}.
\label{eq:two_third}
\end{equation}

Again, by exploiting the isotropic relations in equation (\ref{eq:lontra_link_en}), it is possible to obtain similar scaling relations for the other second-order structure functions:
\begin{equation}
\langle \delta q^2 \rangle \sim \langle \delta u_{\perp}^2 \rangle \sim \langle \varepsilon \rangle^{2/3} r^{2/3},
\label{eq:two_third_2}
\end{equation}
and by further exploiting equation (\ref{eq:lontra_link_en}), it is possible to obtain the ratio between the transverse and longitudinal contributions in the inertial subrange:
\begin{equation}
        \langle \delta u_{\perp}^2 \rangle = \frac{4}{3} \langle \delta u_{||}^2 \rangle.
\label{eq:intertial_scaling_transv}
\end{equation}

{Figure \ref{fig:ener_struct_fun} shows that the present DNS agree with both the theoretical viscous and inertial scalings (equations (\ref{eq:viscous_scaling}) and (\ref{eq:two_third_2})) and with the theoretical ratios between the longitudinal and transverse structure functions in the viscous and inertial subranges (equations (\ref{eq:visc_en_ratio}) and (\ref{eq:intertial_scaling_transv}), respectively). Hence, the present DNS shows to be in accordance with the theoretical results from the literature, attesting that the present data are representative of homogeneous isotropic turbulence.}


Finally, in order to provide a complete reference for the comparison with the analysis of enstrophy described in the core of the present work, we report in figure \ref{fig:ener_flux} the inertial energy flux decomposed in its longitudinal and transverse contributions $\langle \delta q^2 \delta u_r \rangle = \langle \delta u_{||}^3 \rangle + 2 \langle \delta u_{\perp}^2 \delta u_{||} \rangle$, respectively.

Exact relations for these fluxes in the inertial energy subrange ($\eta \ll r \ll L_{11}$) can be derived as follows. Since in the inertial energy subrange viscous effects are negligible, equation (\ref{eq:iso_flux_gke}) reduces to,
\begin{equation}
        \langle \delta q^2 \delta u_{||} \rangle = -\frac{4}{3} \langle \varepsilon \rangle r,
\label{eq:inertial_scaling_flux}
\end{equation}
From this equation, together with equation (\ref{eq:four_fifth}) and the definition $\langle \delta q^2 \delta u_{||} \rangle = \langle \delta u_{||}^3 \rangle + 2 \langle \delta u_{\perp}^2 \delta u_{||} \rangle$ we can obtain an exact relation between the inertial energy flux contributions (in the inertial subrange):
\begin{equation}
        \langle \delta u_{\perp}^2 \delta u_{||} \rangle / \langle \delta u_{||}^3 \rangle = \frac{1}{3}.
\label{eq:tralon_ratio_energy_inert}
\end{equation}

On the other hand, in the viscous subrange, the inertial fluxes of energy may be estimated as $\langle \delta q^2 \delta u_{||} \rangle \sim r^3$. Hence, using this proportionality, together with the identity:
\begin{equation}
\langle \delta q^2 \delta u_{||} \rangle = \frac{1}{3 r^3} \frac{d}{d r} \left(r^4 \langle \delta u_{||}^3 \rangle \right)
\end{equation}
and $\langle \delta q^2 \delta u_{||} \rangle = \langle \delta u_{||}^3 \rangle + 2 \langle \delta u_{\perp}^2 \delta u_{||} \rangle$ it is possible to obtain another exact relation between the inertial energy flux contributions in the viscous subrange:
\begin{equation}
\langle \delta u_{\perp}^2 \delta u_{||} \rangle / \langle \delta u_{||}^3 \rangle = \frac{2}{3}.
\label{eq:tralon_ratio_energy_visc}
\end{equation}
Figure \ref{fig:ener_flux} shows that both viscous and inertial scaling laws and ratios are in accordance with the theoretical results, again attesting the robustness of the present DNS data for conducting this study.

\section{Sensitivity and robustness of the correlation coefficients} \label{robustness}

In the present appendix we show the sensitivity and robustness of the correlation coefficients presented in section \ref{pressure}. We consider three aspects of the analysis. First, we examine the effect of centering the normalized cross-moments, i.e. comparing the results obtained using the original (uncentered) quantities with those obtained after subtracting their respective conditional averages.

In the main analysis, we use the uncentered quantities because we are interested in the net effect of the interscale transfer. However, as a robustness check, we consider here the corresponding centered quantities, which isolate the correlations associated with fluctuations around the conditional mean values. In particular, we define the fluctuating quantities as $T'_{|| / \perp} = T_{|| / \perp} - \langle T_{|| / \perp} \rangle_{10^{+/-}}$, where the conditional average is taken separately for positive and negative events: the $\langle \cdot \rangle_{10^+}$ operator is used when conditioning on positive events, while the $\langle \cdot \rangle_{10^-}$ operator is used when conditioning on negative events. The quantities $T'_{\omega_{|| / \perp}} = T_{\omega_{|| / \perp}} - \langle T_{\omega_{|| / \perp}} \rangle_{10^{+/-}}$ and $T'_{p ||} = T_{p ||} - \langle T_{p ||} \rangle_{10^{+/-}}$ are defined analogously. In both cases, the $\langle \cdot \rangle_{10^{+/-}}$ operator refers to a conditional average on the energy interscale transport $T_{|| / \perp}$.

\begin{figure}[t!]
\centering
\includegraphics[width=0.45\linewidth]{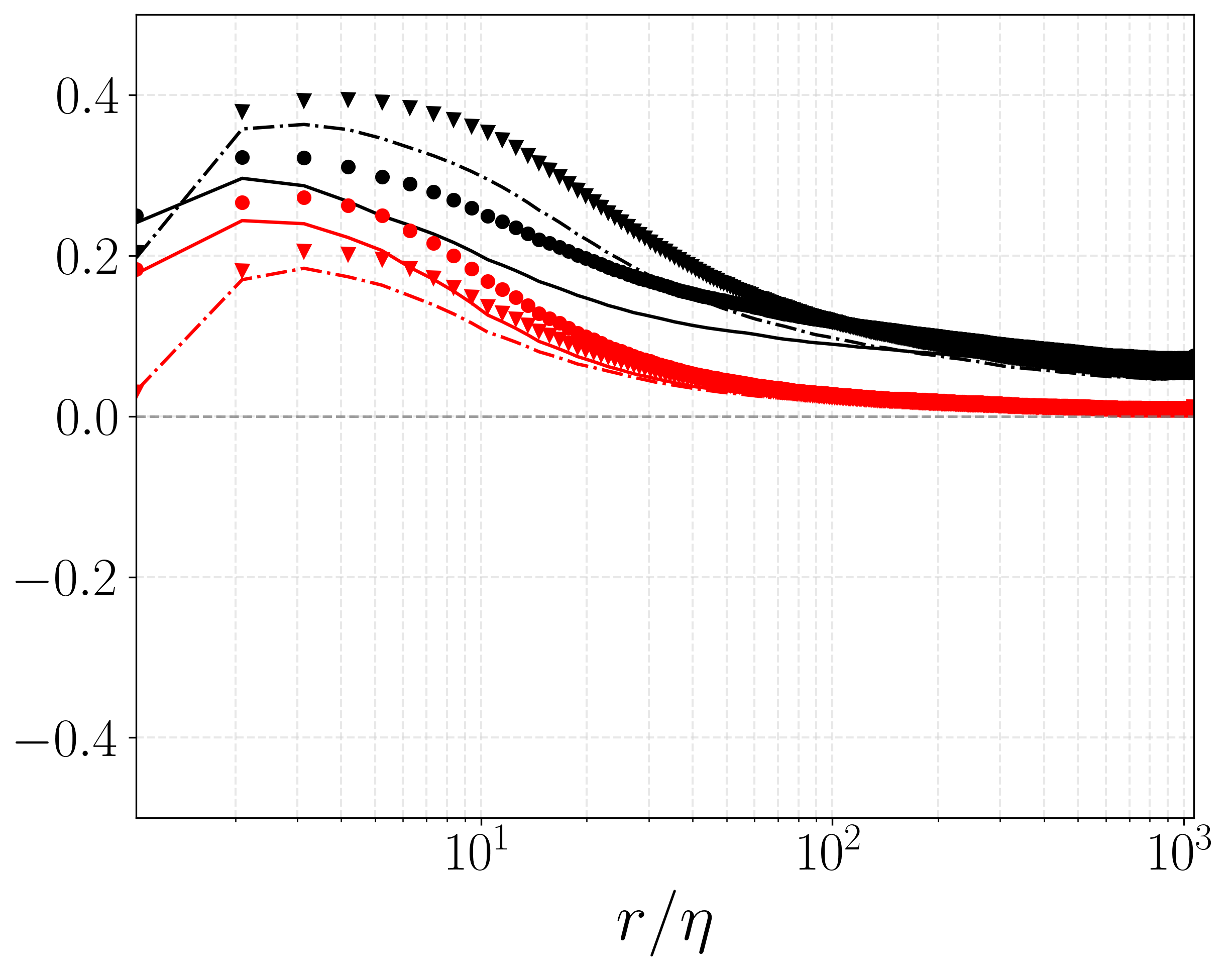}
\includegraphics[width=0.45\linewidth]{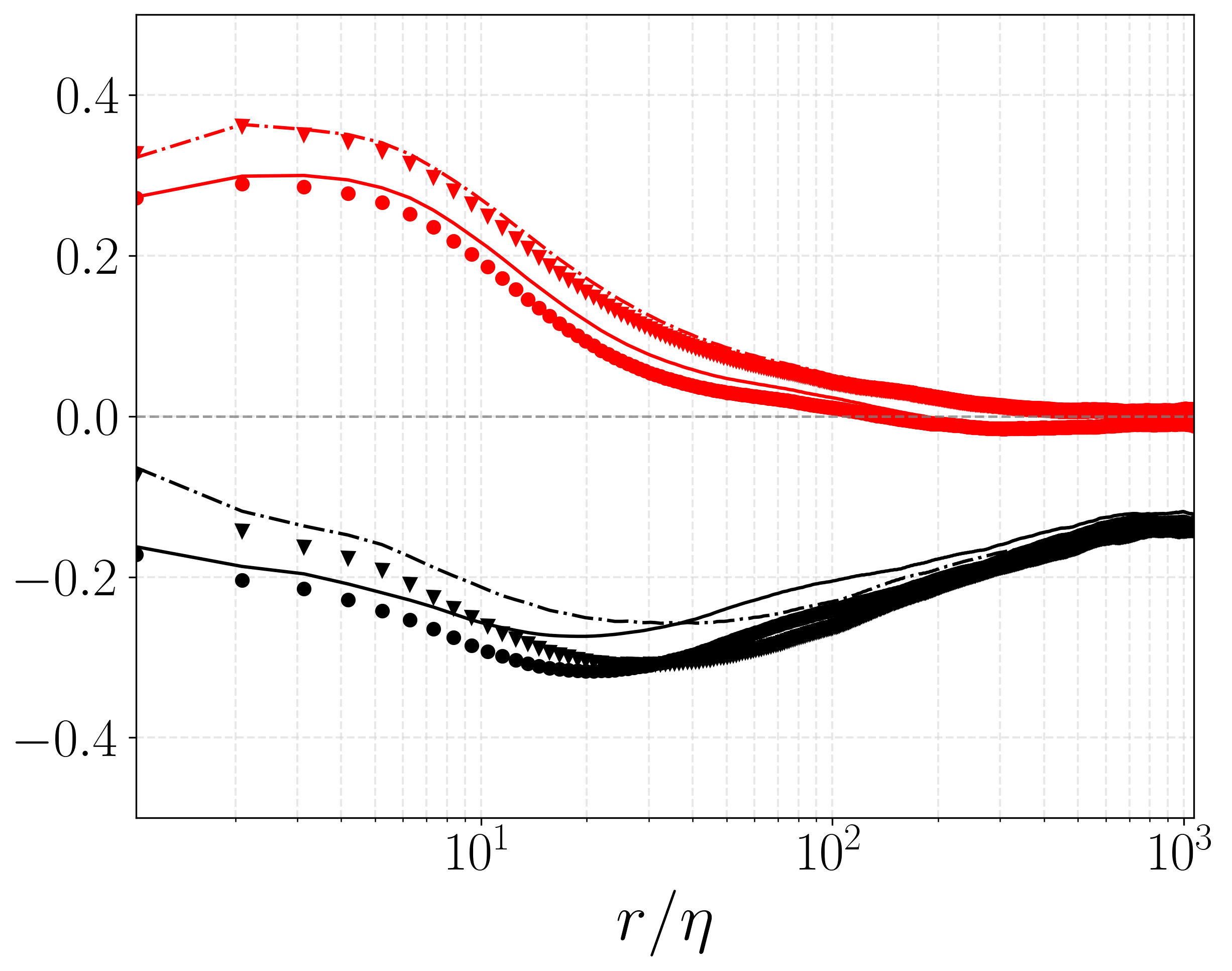}
        \put(-455,185){(a)}
        \put(0,185){(b)}
        \caption{Comparison between uncentered and centered correlation coefficients. Black and red lines denote the uncentered correlations involving $T_{||}$ and $T_{\perp}$, respectively, while black and red symbols denote the corresponding centered correlations involving $T'_{||}$ and $T'_{\perp}$. Solid lines and squares correspond to positive energy transport events, while dash-dotted lines and triangles correspond to negative energy transport events. (a) Correlations with $T_{\omega_{||/\perp}}$ and $T'_{\omega_{||/\perp}}$. (b) Correlations with $T_{p||}$ and $T'_{p||}$. All quantities are conditioned on intense energy interscale transport events using thresholds of $10\langle T_{|| / \perp} \rangle$.}
\label{fig:centered}
\end{figure}

The results, shown in figure \ref{fig:centered}, exhibit similar trends and do not lead to any significant qualitative differences or changes in the conclusions drawn in section \ref{pressure}.

\begin{figure}[b!]
\centering
\includegraphics[width=0.45\linewidth]{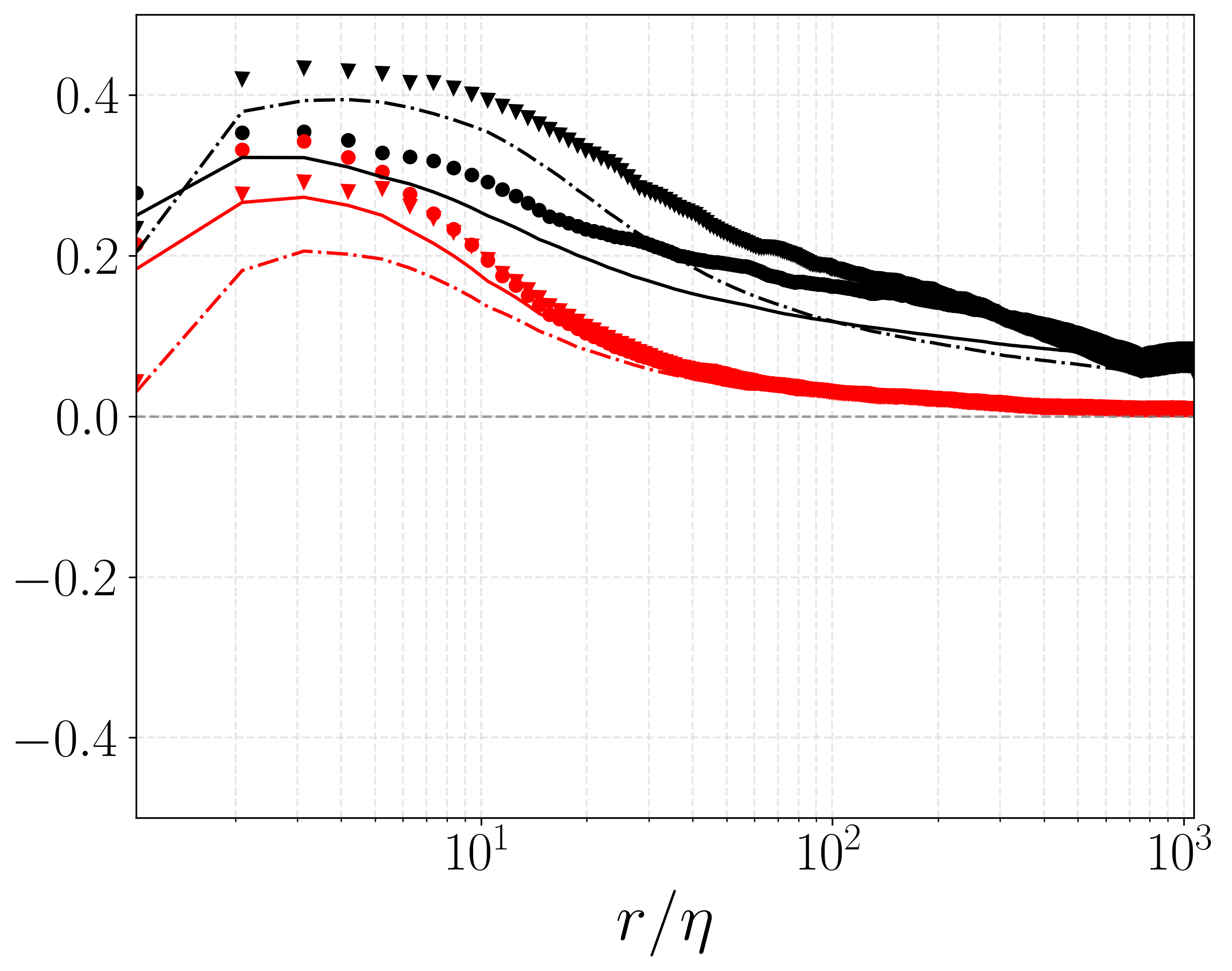}
\includegraphics[width=0.45\linewidth]{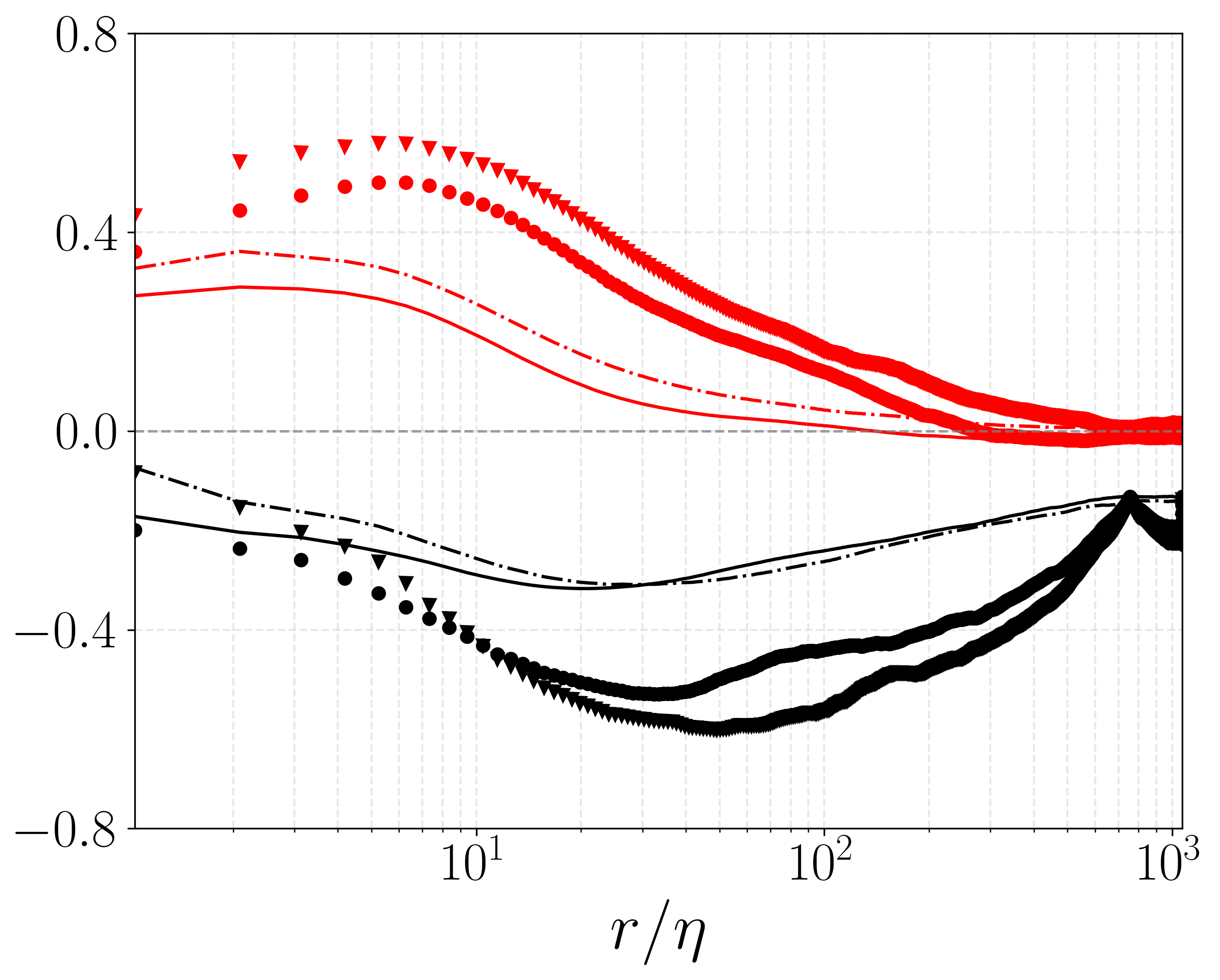}
        \put(-455,185){(a)}
        \put(0,185){(b)}
        \caption{Effect of the threshold on the correlation coefficients. In both panels, black and red curves denote the correlations involving $T_{||}$ and $T_{\perp}$, respectively. Solid and dash-dotted lines correspond to positive and negative energy transport events, respectively, with a threshold of $10\langle T_{|| / \perp}\rangle$. The corresponding correlations obtained with a threshold of $1000\langle T_{|| / \perp}\rangle$ are shown by squares for positive energy interscale transport events and triangles for negative events. (a) Correlations with $T_{\omega_{||/\perp}}$. (b) Correlations with $T_{p||}$.}
\label{fig:1000}
\end{figure}

As a second test of the robustness of the results, we investigate the sensitivity of the correlation coefficients to the choice of threshold used to identify the ``intense'' energy interscale transport events. In particular, we tested thresholds ranging from one to 1000 times the average value. In figure \ref{fig:1000} we compare the results presented in section \ref{pressure} ($10 \langle T_{|| / \perp} \rangle$) with those obtained using the most extreme threshold tested, namely $1000 \langle T_{|| / \perp} \rangle$. The use of a higher threshold resulted in higher correlation coefficients, while preserving the qualitative trends of the curves, thus leaving the considerations made in section \ref{pressure} unchanged.

\begin{figure}[t!]
\centering
\includegraphics[width=0.45\linewidth]{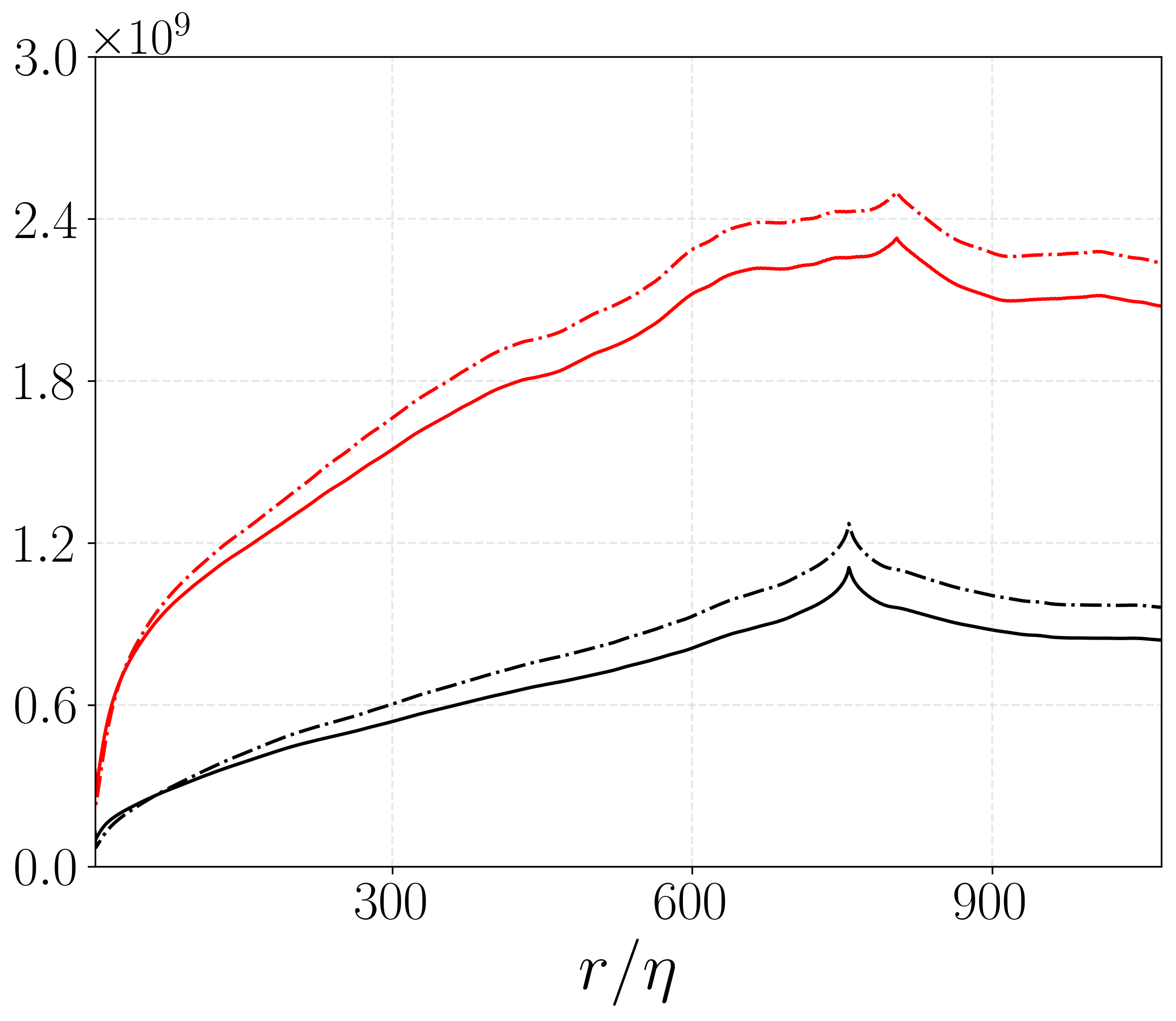}
\includegraphics[width=0.45\linewidth]{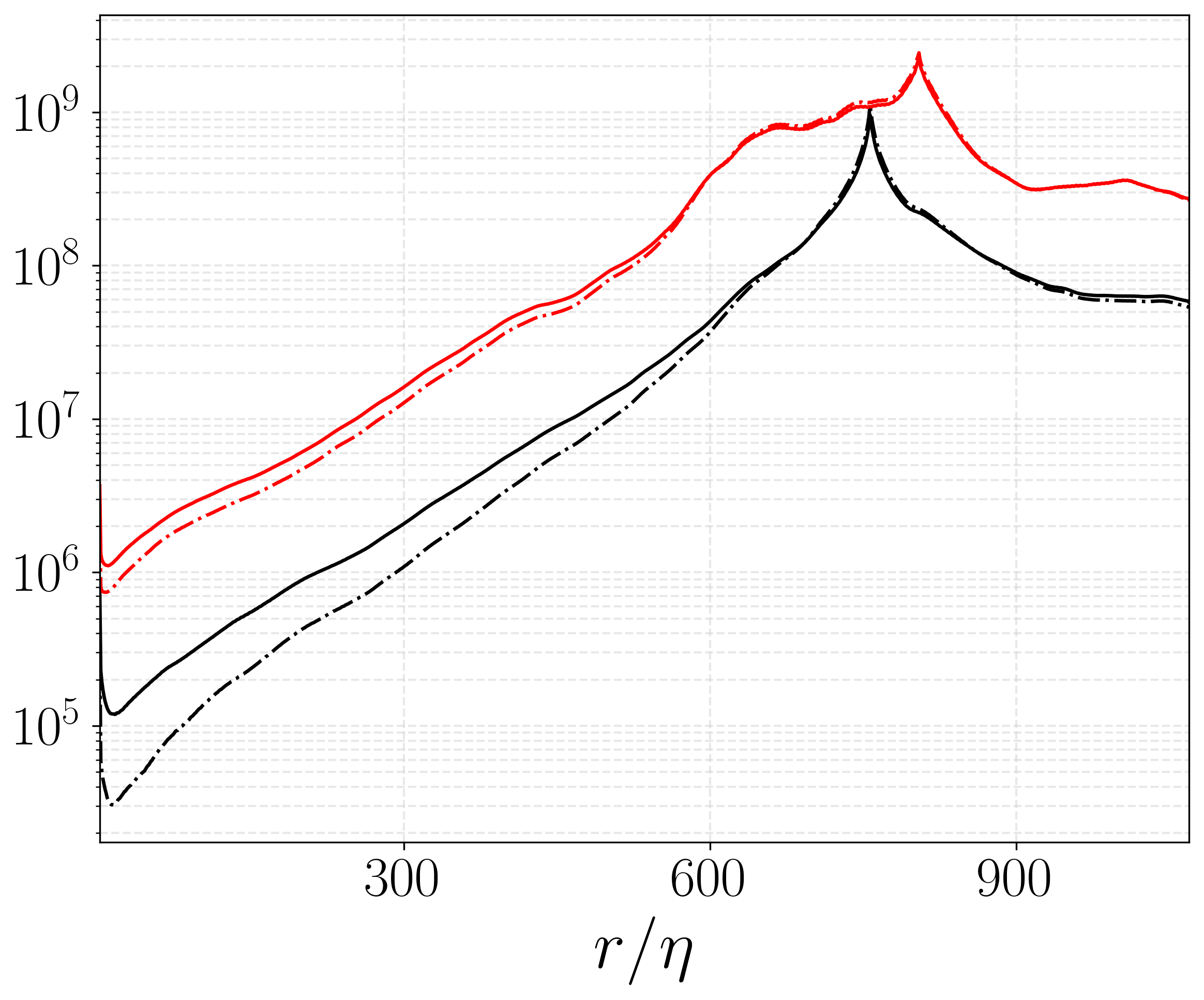}
        \put(-465,198){(a)}
        \put(4,198){(b)}
        \caption{Number of samples used to compute the correlation coefficients for thresholds of (a) $10\langle T_{|| / \perp}\rangle$ and (b) $1000\langle T_{|| / \perp}\rangle$. Black and red lines denote conditioning on longitudinal ($T_{||}$) and transverse ($T_{\perp}$) energy transport, respectively; solid and dash-dotted lines correspond to positive and negative events, respectively.}
\label{fig:samp_count}
\end{figure}

We conclude this appendix by assessing the convergence of the curves presented above. In particular, we show the number of samples collected for the two thresholds considered here, namely $10$ and $1000$ times the average values $\langle T_{|| / \perp} \rangle$, in figure \ref{fig:samp_count}(a) and (b), respectively. For this analysis, we collected one sample every four grid points. Although this reduces the total number of samples, the larger spatial separation between consecutive samples makes them less correlated, thereby improving the statistical convergence for a given number of samples. The number of samples varies with the scale considered and exhibits a peak at the scale where the corresponding mean interscale energy transport vanishes, i.e. $\langle T_{||} \rangle = 0$ or $\langle T_{\perp} \rangle = 0$.

\begin{figure}[b!]
\centering
\includegraphics[width=0.45\linewidth]{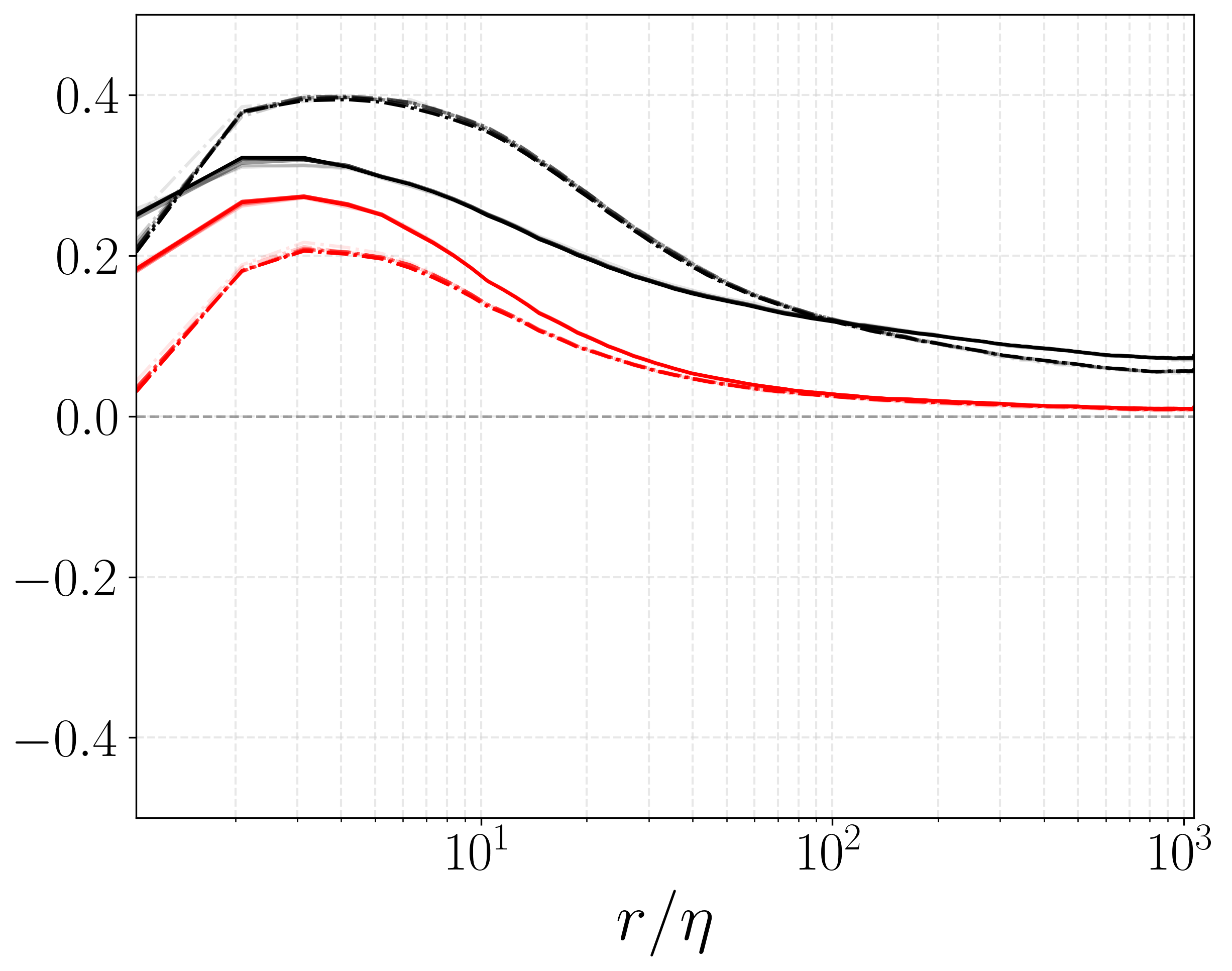}
\includegraphics[width=0.45\linewidth]{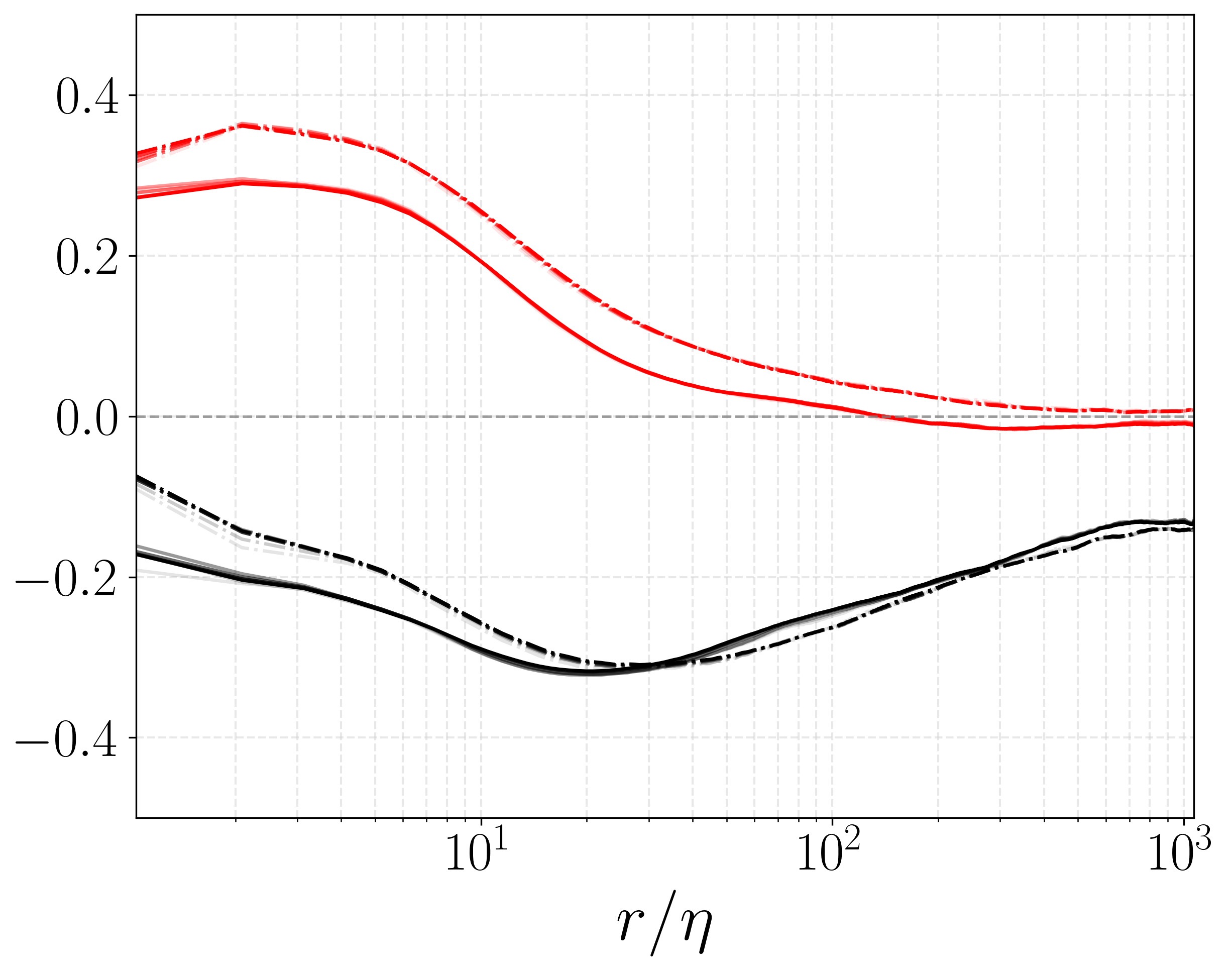}
        \put(-455,175){(a)}
        \put(0,175){(b)}
        \caption{Snapshot-to-snapshot convergence. Transparency indicates the number of temporal samples, from 1 (most transparent) to 6 (opaque). Black and red lines denote correlations involving $T_{||}$ and $T_{\perp}$, respectively, while solid and dash-dotted lines correspond to positive and negative energy transport events, respectively. (a) Correlations with $T_{\omega_{||/\perp}}$. (b) Correlations with $T_{p||}$. All quantities are conditioned on intense energy interscale transport events using thresholds of $10\langle T_{|| / \perp} \rangle$.}
\label{fig:snap_to_snap}
\end{figure}

Finally, the snapshot to snapshot convergence of the results presented in section \ref{pressure}, reported in figure \ref{fig:snap_to_snap}, shows that the correlation coefficients are well-converged. Despite the relatively small number of temporal snapshots used, the correlation coefficients converge more rapidly than the individual variances and covariances entering their definition, likely as a result of the normalization involved.

\bibliography{bibliography}

\end{document}